\documentclass[notitlepage,aps,showpacs,prd]{revtex4-1}

\usepackage[ansinew]{inputenc}
\usepackage{amsmath}
\usepackage{amsfonts}
\usepackage{amssymb}
\usepackage{mathrsfs}
\usepackage{color}
\usepackage{graphicx}
\usepackage{pstricks}
\usepackage{subfigure}

\begin{document}

\title{Dynamics of test particles in thin-shell wormhole spacetimes}

\author{Valeria Diemer\footnote{n\'{e}e Kagramanova} and Elena Smolarek} 
\affiliation{
Department of Physics, Carl von Ossietzky University, 26111 Oldenburg, Germany
}

\date{\today}

\begin{abstract}

Geodesic motion in traversable Schwarzschild and Kerr thin-shell wormholes constructed by the cut-and-paste method introduced by Visser~\cite{Visser2,Visserbook} is studied. 
The orbits are calculated exactly in terms of elliptic functions
and visualized with the help of embedding diagrams. 

\end{abstract}

\maketitle

\section{Introduction}

The idea to travel in the twinkling of an eye from one region of the Universe to another, no matter how far, or even to travel in time has been fascinating people since many years. Wormholes are believed to make such dreams possible. The first solution providing such a saught for connection between distant locations was discovered by Einstein and Rosen~\cite{EinRose}, the Einstein-Rosen bridge. Wheeler gave it the name wormhole~\cite{Wheeler1962}. But unfortunately this first wormhole was not traversable~\cite{FullWheel}. 

For Sagan's novel ``Contact'' Morris and Thorne devised a wormhole which could in principle be traversed by humans~\cite{MorrisThorne}. Visser~\cite{Visser1,Visser2} provided further examples of traversable wormholes. In particular, he constructed such wormholes by surgically manipulating the Schwarzschild spacetime so that no event horizon was present. In both cases, the price to pay is the need for an exotic form of matter that must be present at the throat of such wormholes. Moreover, such wormholes should be stable if they are traversable. Visser~\cite{Visser2} stressed that certain equations of state would lead to stable wormholes. He shows that a traveler can prevent from interacting with the exotic matter and feels no tidal forces during a journey through a thin-shell wormhole. A detailed study of the stability of these traversable wormholes against radial perturbations can be found in~\cite{PoissonVisser1995} where the conditions on the wormhole mass, throat radius and equation of state parameter defined by a particular model of the exotic matter are derived. Linearized radial stability of charged thin-shell wormholes is studied for example in~\cite{Eiroa5} and generalization to the presence of a cosmological constant is considered in~\cite{LoboCrawford2004}.  Stability of the general thin-shell wormholes with spherical symmetry and its dependence on a suitable exotic material on the wormhole throat are investigated in~\cite{Eiroa6,LoboCrawford2005,GarciaLoboVisser2012}. 

The material comprising the throat of the wormhole was addressed in many scientific works and is still an active field of research. For example, a phantom scalar field was discussed in~\cite{Sushkov1,Lobo1,Zaslavskii,Kuhfittig1,Rahaman06},  `tachyon matter' as a source term in the field equations with a positive cosmological constant was studied in~\cite{DasKar}, or the wormhole geometry with a Chaplygin gas in the exotic equation of state was explored in~\cite{Lobo2,Eiroa1,Eiroa2}. An interesting discussion of a perfect fluid and an anisotropic fluid as candidates for the exotic matter was given in~\cite{KashSush2011}. Since the metric is continuous on a throat but not its first derivatives, the energy conditions on the matter supporting the wormhole can be derived by considering the Riemann curvature on both sides of the throat. Thus, from the conditions on the surface energy density, surface pressures and angular momentum density follow that a perfect fluid can only support a spherically symmetric thin-shell wormhole~\cite{Visser1,Visser2}, while an anisotropic fluid can maintain the geometry of a rotating thin-shell wormhole~\cite{KashSush2011}. But the matter in both cases has negative surface energy density violating the weak energy conditions. Estimates on the amount of exotic matter necessary for the traversability were given in~\cite{Kuhfittig3}. A combined model comprising ordinary and quintessential matter can support a traversable wormhole in  Einstein-Maxwell gravity as shown in~\cite{Kuhfittig2}. Wormholes in the framework of the Brans–Dicke gravity were constructed in~\cite{Richarte3}. Further references can be found in the book of Visser~\cite{Visserbook} and in the overview by Lobo~\cite{Lobobook}. The appearance of wormholes in string theory was investigated in~\cite{string1,string2,Eiroa3,Eiroa4,Richarte1,Richarte2}.  

Stable spherical wormholes which do not need any form of exotic matter for their existence were recently obtained numerically in dilatonic Einstein-Gauss-Bonnet theory in four spacetime dimensions~\cite{KKK2011,KKK2012}. Besides their stability the authors studied their domain of existence, they investigated their geodesics, determining the possible types of trajectories, and performed a study of the acceleration and tidal forces that a traveler crossing such a wormhole would feel. Further numerically obtained configurations of wormholes were explored in~\cite{Kirg1,Kirg2,Kirg3}, where wormholes were filled by a perfect fluid (ordinary matter) and a phantom scalar field. This model was applied to describe stars as well as neutron stars with a nontrivial topology. Traversable wormholes without violation of energy conditions in the geometries of charged shells are constructed analytically in~\cite{Schein96}. However, these wormhole spacetimes have closed timelike curves.

A comprehensive investigation of the geodesics in the Morris-Thorne wormhole spacetime was carried out in~\cite{Mueller2008}, where for visualization of the trajectories embedding diagrams were constructed. Moreover, gravitational lensing and illumination calculations were addressed. Further properties of the propagation of particles and fields in static spherically symmetric wormhole spacetimes were studied in~\cite{SarbachZannias}. Some aspects of the electromagnetic fields and charged particle motion around slowly rotating magnetized wormholes are discussed in~\cite{AbduAhm}.

Rotating wormholes are a natural generalization of the initially studied, static spherically symmetric wormholes, that are interesting from a physical point of view. For instance, one can construct a thin-shell rotating wormhole~\cite{KashSush2011} by the same cut-and-paste method~\cite{Visser2} used for the thin-shell static wormhole. On the other hand, generalizations of the Morris-Thorne wormhole~\cite{MorrisThorne} to rotating wormholes were presented by Teo~\cite{Teo} and Kuhfittig~\cite{Kuhfittig4}. It was shown that these wormholes are traversable (tidal forces are not larger than on Earth) and due to the rotation and a possible ergoregion (for not too wide a throat) an extraction of energy by the Penrose process is possible. Slowly rotating wormholes with a phantom scalar field were studied in~\cite{KashSush2,MatosNunez}. The electromagnetic field generation around rotating wormholes surrounded by charged particles was studied in~\cite{Jamil}. In~\cite{Matos2010} a class of rotating and magnetized wormholes with Einstein-–Maxwell and phantom fields was explored.

The conversion of a wormhole into a time machine, the problems appearing thereby and accompanying physical effects were discussed in~\cite{timemachine1,timemachine2,timemachine3}. The associated problems of causality violations were addressed in~\cite{KimThorne,Hawking}.

In this article we study the motion of massive and massless particles in thin-shell Schwarzschild and Kerr wormhole spacetimes, constructed by the cut-and-paste method~\cite{Visser2,KashSush2011}. Thus the paper contains two main sections. Section~\ref{sec:schw} explores the dynamics of null and timelike geodesics for static spherically symmetric thin-shell wormholes and Section~\ref{sec:kerr} - the geodesics for rotating thin-shell wormholes. We also study the influence of the ergoregion, existing for sufficiently thin throats, on the properties of the orbits. The trajectories are given by exact analytical solutions of the geodesic equations in terms of the Jacobi and Weierstrass elliptic functions.

\section{Schwarzschild wormhole}\label{sec:schw}

\subsection{The geodesic equation}

We start with the Schwarzschild metric in the form (see e.g.~\cite{Chandrasekhar83,Hartle03,Schutz85})
\begin{equation}
ds^2 = - \left(1 - \frac{1}{r} \right) dt^2 + \left(1 - \frac{1}{r} \right)^{-1} dr^2 + r^2 \left( \sin^2\vartheta d\varphi^2 + d\vartheta^2 \right) \ ,  \label{eq:schw_metric}
\end{equation}
where the radial coordinate $r$ is normalized to $2M$ and is dimensionless.

For free particles moving along geodesics in the equatorial plane the Lagrangian $L$ takes the form~\cite{Chandrasekhar83,Shapiro}
\begin{equation}
2 \mathcal{L} = - \left(1 - \frac{1}{r} \right) \dot{t}^2 + \left(1 - \frac{1}{r} \right)^{-1} \dot{r}^2 + r^2 \dot{\varphi}^2 \,\,\,\, {\rm with} \,\,\,\, 2 \mathcal{L} = - \delta \ , \label{eq:schw_L}
\end{equation}
where $\dot{x}^\alpha$ denotes the differentiation with respect to the affine parameter $\lambda$ and $\delta=0$ for null and $\delta=1$ for time-like geodesics. 

With the conserved energy $E= - \frac{\partial \mathcal{L} }{\partial \dot{t}} = \left(1 - \frac{1}{r} \right) \dot{t}$ and conserved angular momentum $L=\frac{\partial \mathcal{L}}{\partial \dot{\varphi}} = r^2 \dot{\varphi}$, a free test particle moves along a geodesic defined by the diffenrential equation
\begin{equation}
\left( \frac{dr}{d\varphi} \right)^2 = \frac{r^4}{L^2}\left( E^2 - \left(\delta+\frac{L^2}{r^2}\right)\left(1-\frac{1}{r}\right) \right) \ ,  \label{eq:schw}
\end{equation}
where the quantities $E$ and $L$ are dimensionless.  

To solve the differential equation~\eqref{eq:schw} we refer to the theory of elliptic functions~\cite{Markush} and make few substitutions in order to reduce the problem to the standard form. A first substitution $r=u^{-1}$ reduces~\eqref{eq:schw} to the form
\begin{equation}
\left( \frac{du}{d\varphi} \right)^2 = u^3 - u^2 + \frac{\delta}{L^2}u + \frac{E^2-\delta}{L^2} \ . \label{eq:schw2}
\end{equation}

The differential~\eqref{eq:schw2} is elliptic of the first kind and can be integrated by Jacobi elliptic functions. To reduce it to the standard form we make two transformations. The first one $u=4v+\frac{1}{3}$ transforms~\eqref{eq:schw2} into the Weierstrass form
\begin{equation}
\left( \frac{dv}{d\varphi} \right)^2 = 4 \prod_{i=1}^3{(v-v_i)} = 4 v^3 - g_2 v - g_3 \ , \label{eq:schw3} 
\end{equation}
where $v_i$, $i=1,2,3$, are the roots of the cubic polynomial in~\eqref{eq:schw3} and satisfy the condition $\sum_{i=1}^3{v_i}=0$. $g_3=4v_1v_2v_3$ and $g_2=4v_3^2-4v_1v_2$.

The second transformation $v=(v_2-v_1)s^{-2}+v_1$ turns~\eqref{eq:schw3} into
\begin{equation}
\frac{1}{v_2-v_1}\left( \frac{ds}{d\varphi} \right)^2 = (1-s^2)(1-m^2s^2) \ , \label{eq:schw4} 
\end{equation}
where $m^2=(v_3-v_1)(v_2-v_1)^{-1}$, and $m$ is the modulus of the Jacobian elliptic functions~\cite{Markush}.

With Jacobi's elliptic function $sn(x)$ we find from~\eqref{eq:schw4} $s$ as a function of $\varphi$:
\begin{equation}
s = sn\Bigl( (\varphi-\varphi^\prime)\sqrt{v_2-v_1} \Bigr) \ , \label{eq:schw5} 
\end{equation}
where 
\begin{equation}
\varphi^\prime = \varphi_0 + \frac{1}{\sqrt{v_2-v_1}} \int^0_{s_0} \frac{ds}{ \sqrt{ (1 - s^2)(1-m^2 s^2) } } \,
\end{equation}
is a constant which can be expressed in terms of the periods of $sn(x)$, and $\varphi_0$ is the initial value of $\varphi$.

We then get the function $r(\varphi)$~\cite{Chandrasekhar83,Markush}:
\begin{equation}
r = \left( \frac{4(v_2-v_1)}{sn^2\Bigl( (\varphi-\varphi^\prime)\sqrt{v_2-v_1} \Bigr)} + 4v_1 + \frac{1}{3} \right)^{-1} \ . \label{eq:schw6} 
\end{equation}

\subsection{Construction of the wormhole and embedding}\label{sec:schw_emb}

To obtain the Schwarzschild thin-shell wormhole we follow the `surgery'-method of Visser~\cite{Visser1,Visser2}. This method was also applied by Kashgarin and  Sushkov~\cite{KashSush2011} for the constuction of the Kerr thin-shell wormhole (see Sec.~\ref{sec:kerr_emb}). The procedure is as follows: we cut the Schwarzschild spacetime outside the event horizon $h=1$ at some value of the radial coordinate $b_0$, where $b_0$ is now the throat parameter and $b_0 > h$. Gluing two copies of the asymptotically flat region $r\geq b_0$, which are then connected by a wormhole with the throat hypersurface $\Sigma$, we obtain a geodesically complete manifold. The coordinate $|l| = |r-b_0|$ such that $l\in(-\infty,\infty)$ then covers the whole spacetime (the upper and the lower universe). The throat of the wormhole is located at $l=0$.

To visualize the topology of the wormhole we consider a two-dimensional hypersurface $(t=const, \vartheta=\pi/2)$~\cite{Mueller2008}. Its inner geometry yields
\begin{eqnarray}
ds_0 &=& \left(1 - \frac{1}{r} \right)^{-1} dr^2 + r^2 d\varphi^2 \label{schw_emb1} \ .
\end{eqnarray}
This two-dimensional hypersurface can be embedded into the Euclidean space given by
\begin{equation}
ds_E = dr^2 + r^2 d\varphi^2 + dz^2
     = \left( 1 + \left(\frac{dz}{dr}\right)^2 \right) dr^2 + r^2 d\varphi^2 \label{schw_emb2} \,
\end{equation}
in the cylindrical coordinates $(r,\varphi,z)$. 

Comparing the coefficients of $dr^2$ in~\eqref{schw_emb1} and~\eqref{schw_emb2} we find that the shape of the embedding diagram in the Euclidean space is given by 
\begin{equation}
z(r)= 2\sqrt{r-1} - 2\sqrt{b_0-1}  \ . \label{schw:emb}
\end{equation}

It is visualized in Fig.\ref{fig:schworb} in coordinates $(x,y,z)$, where $x=r\cos(\varphi)$ and $y=r\sin(\varphi)$. For the embedding diagram we choose the initial value of $r$ to be $b_0$ and $\varphi\in[0,2\pi]$. (In the next section we discuss the orbits in the thin-shell Schwarzschild-wormhole spacetime and show an example for the embedding diagram of an orbit.)

\subsection{Geodesics}\label{sec:schw_geod}

Let us introduce an effective potential in equation~\eqref{eq:schw}
\begin{equation}
\left( \frac{dr}{d\varphi} \right)^2 = \frac{r^4}{L^2}\left( E^2 - V_{\rm{eff}} \right) \label{eq:Veff_in} \ ,
\end{equation}
i.e.~\cite{Chandrasekhar83} 
\begin{equation}
V_{\rm{eff}}=\left( \delta + \frac{L^2}{r^2} \right)\left( 1- \frac{1}{r} \right) \label{eq:Veff} \ ,
\end{equation}
where intersections of the function $E^2$ and $V_{\rm{eff}}$ specify the turning points of the motion.

\begin{figure*}[th!]
\begin{center}
\subfigure[][$L^2=4.5$, $b_0=1.01$]{\label{schw_pot1}\includegraphics[width=8.5cm]{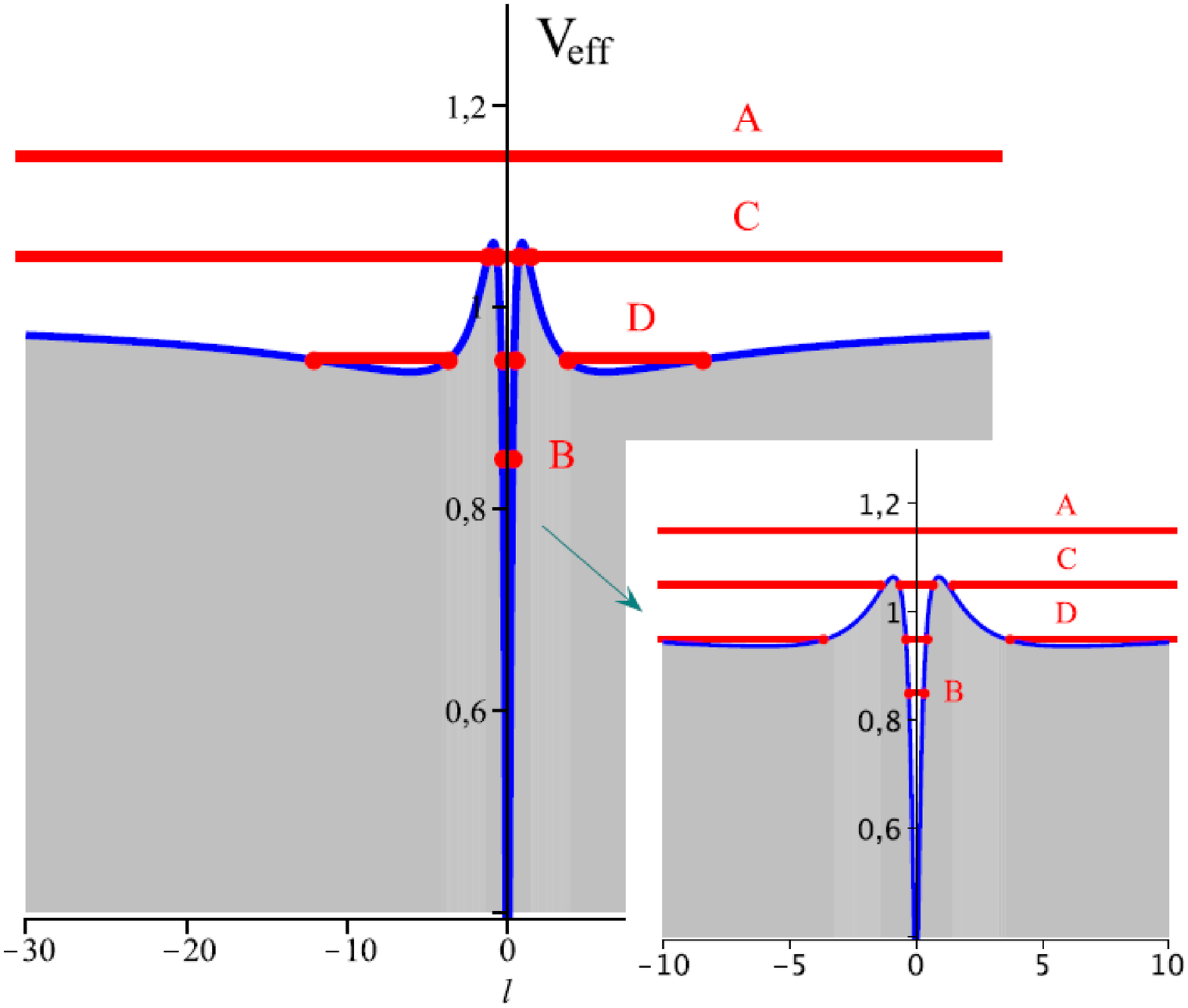}}
\subfigure[][$L^2=4.5$, $b_0=2$]{\label{schw_pot2}\includegraphics[width=7cm]{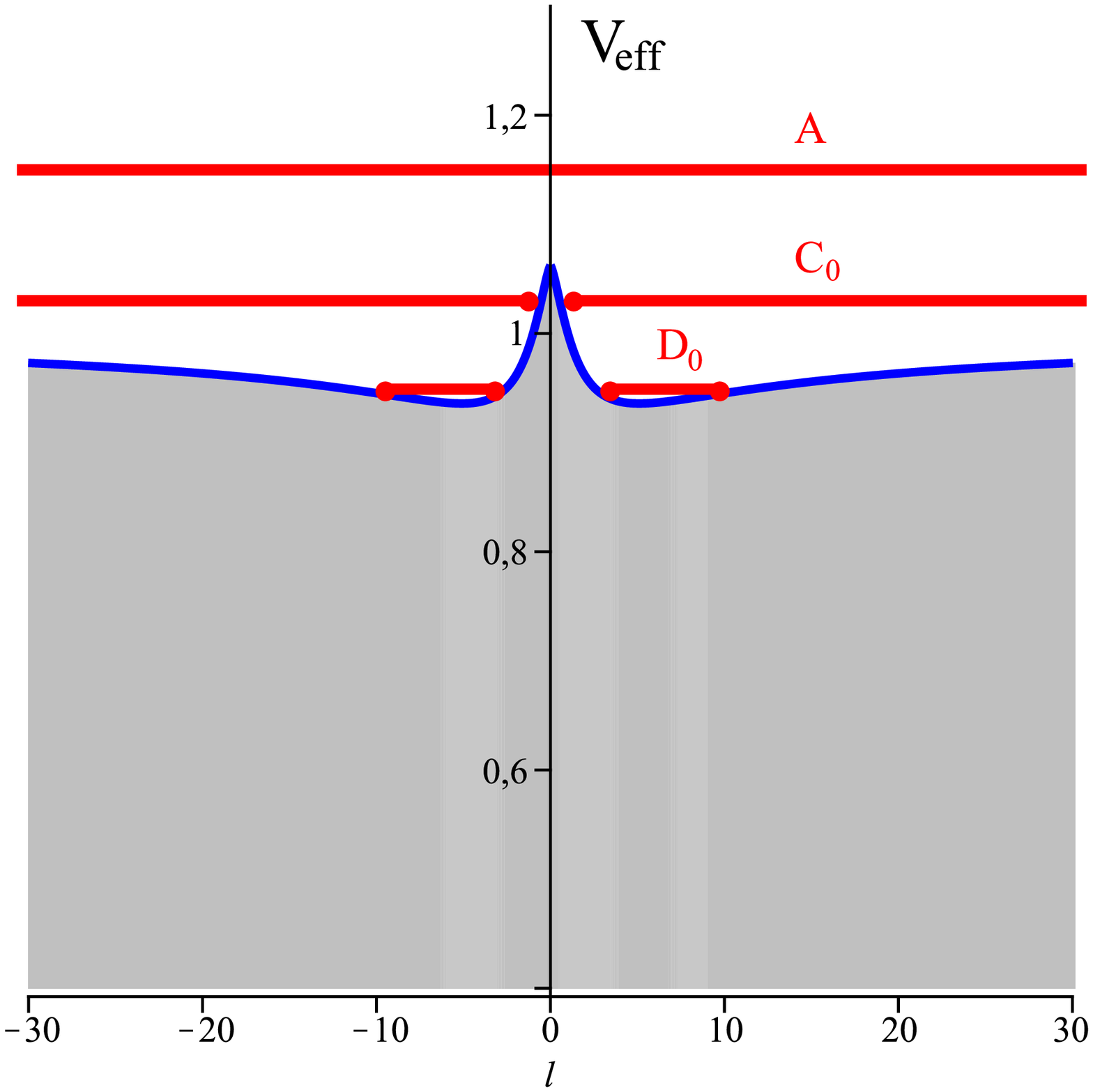}}
\subfigure[][$L^2=4.5$, $b_0=6$]{\label{schw_pot3}\includegraphics[width=7cm]{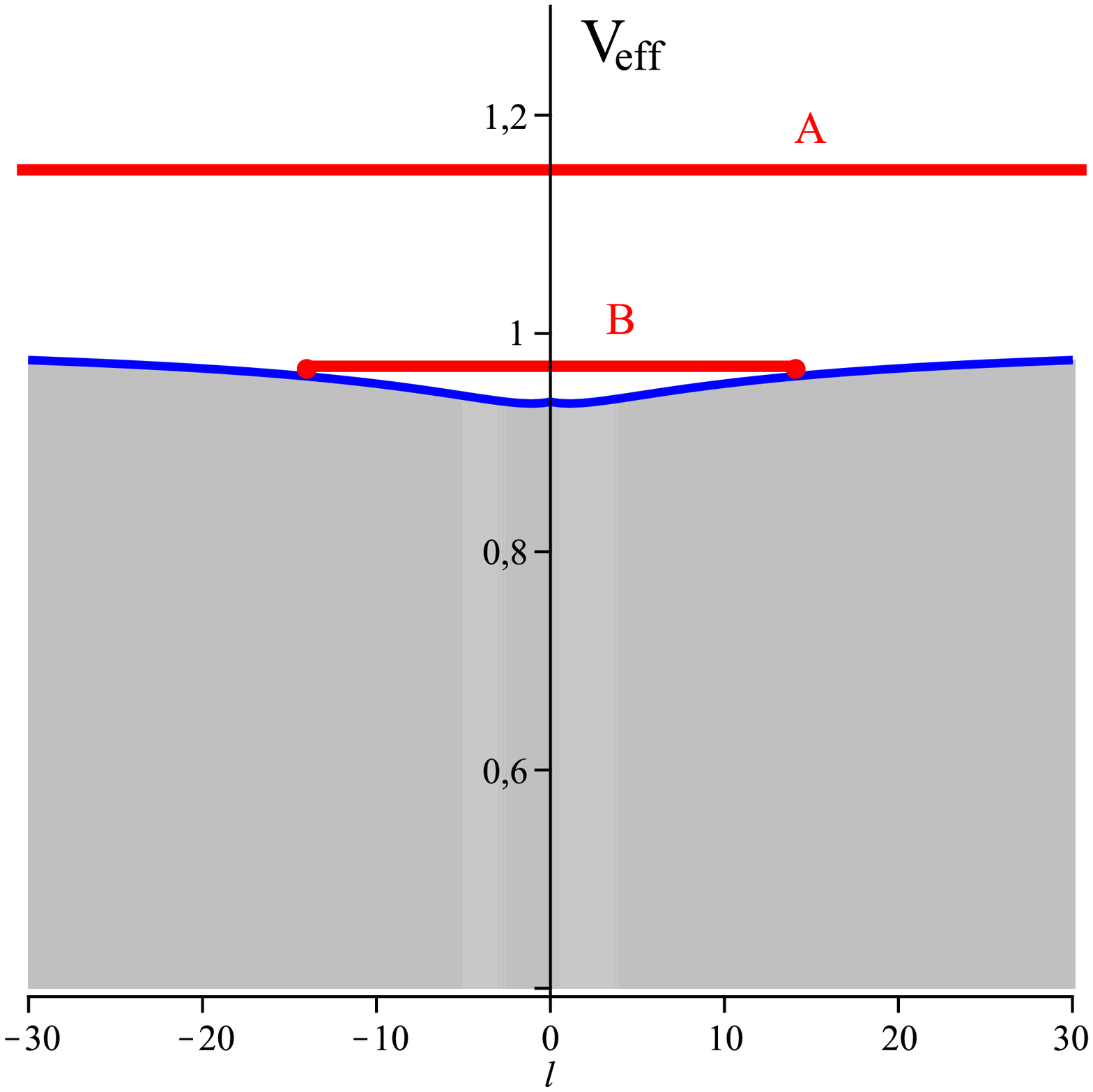}}
\subfigure[][$L^2=9$]{\label{schw_pot4}\includegraphics[width=7cm]{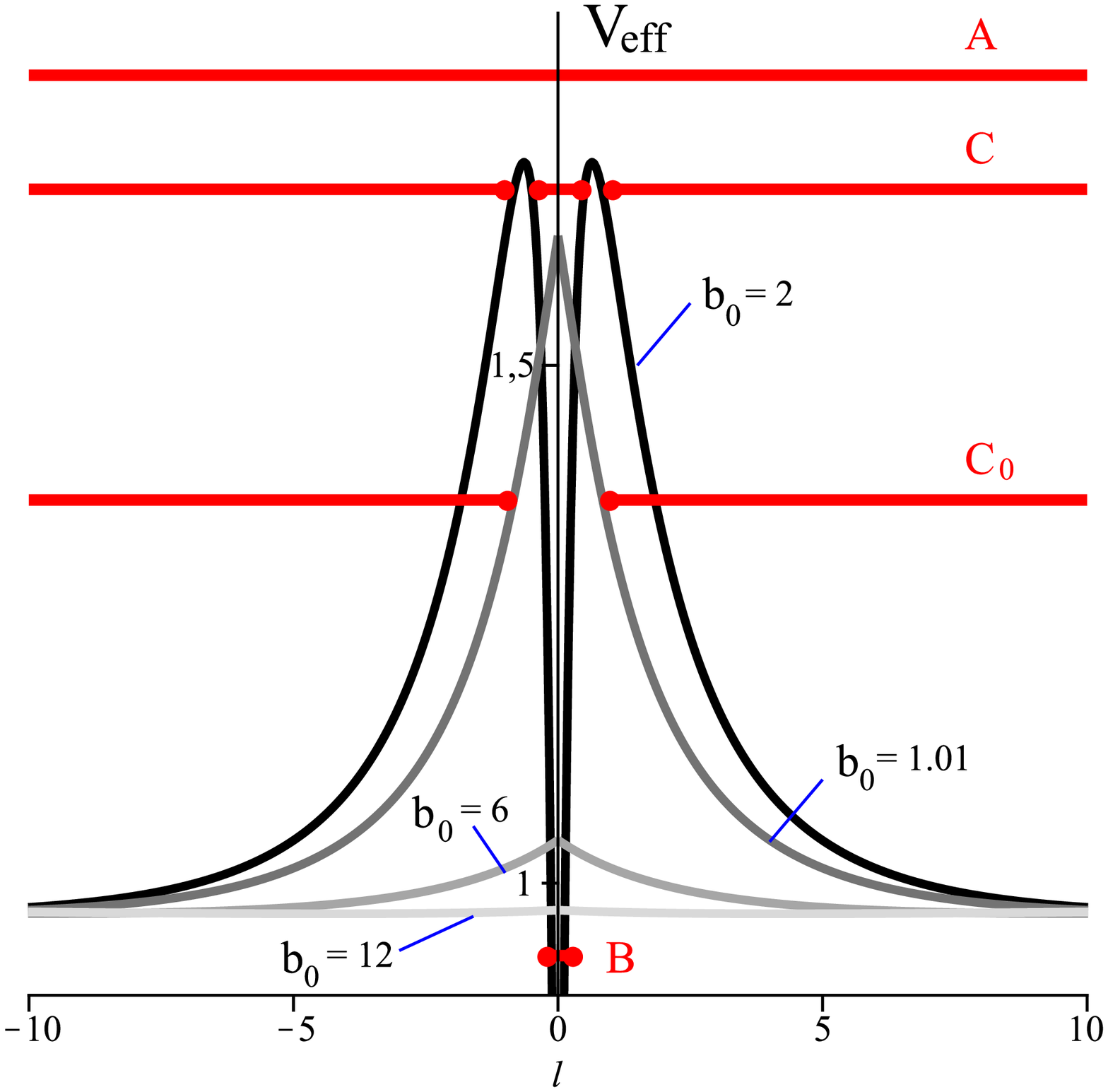}}
\end{center}
\caption{The effective potential (eq.~\eqref{eq:Veff}) in the Schwarzschild wormhole spacetime for $\delta=1$. Filled regions in~\subref{schw_pot1}-\subref{schw_pot3} mean forbidden energy values. Only in~\subref{schw_pot4} we did not color the forbidden regions in order not to overcharge the picture. For a growing throat of the wormhole no planetary bound orbits exist. The orbit of type A corresponds to the two-world escape orbit passing through the upper and lower parts of the Universe. The orbit of type B is a two-world bound orbit. The orbits of type D include planetary bound orbits, and orbits of type C include escape orbits in both parts of the Universe. For details see Sec.~\ref{sec:schw_geod}. \label{fig:schwpot}}
\end{figure*}

Figure~\ref{fig:schwpot} shows examples of the effective potential~\eqref{eq:Veff} for massive test particles with $\delta=1$ for different values of $b_0$. Here the coordinate $l$ was used. The regions which do not satisfy the condition $E^2\geq V_{\rm{eff}}$ are forbidden and colored grey.

The following types of orbits exist in the Schwarzschild-wormhole spacetime for massive test particles:
\begin{itemize}
\item[Type $A$] two-world escape orbit TWE which connect two parts of the Universe. These orbits would be ending at the singularity in usual Schwarzschild spacetime.
\item[Type $B$] two-world bound orbits TWB when a test particle moves on a bound orbit stretching from one into another part of the Universe. This orbit connects the upper and lower parts of the universe similar to the orbit of type A.
\item[Type $C$] this type contains two possible orbits: two-world bound TWB and escape orbits EO. EOs exist in both parts of the Universe. For growing values of the throat parameter $b_0$ this type reduces to $C_0$ containing only EOs. See Fig.\ref{schw_pot2}.
\item[Type $D$] this type includes two-world bound TWB and bound orbits BO. BOs exist in both parts of the Universe. Here again $D$ reduces to type $D_0$ for growing values of the throat parameter $b_0$. See Fig.\ref{schw_pot2}.
\end{itemize}

For large $b_0$-values only orbits of type $B$ and $A$ survive (Fig.\ref{schw_pot3}). For very large values - only type $A$ is possible.

Figure~\ref{schw_pot4} shows a set of potential plots for large $L^2=9$, when no planetary bound orbits are possible (also in the Schwarzschild black hole spacetime), and different values of the throat parameter $b_0$.

\begin{figure*}[th!]
\begin{center}
\subfigure[][$L^2=4.5, E^2=0.949$. TWB.]{\label{schw_orb1}\includegraphics[width=4.5cm]{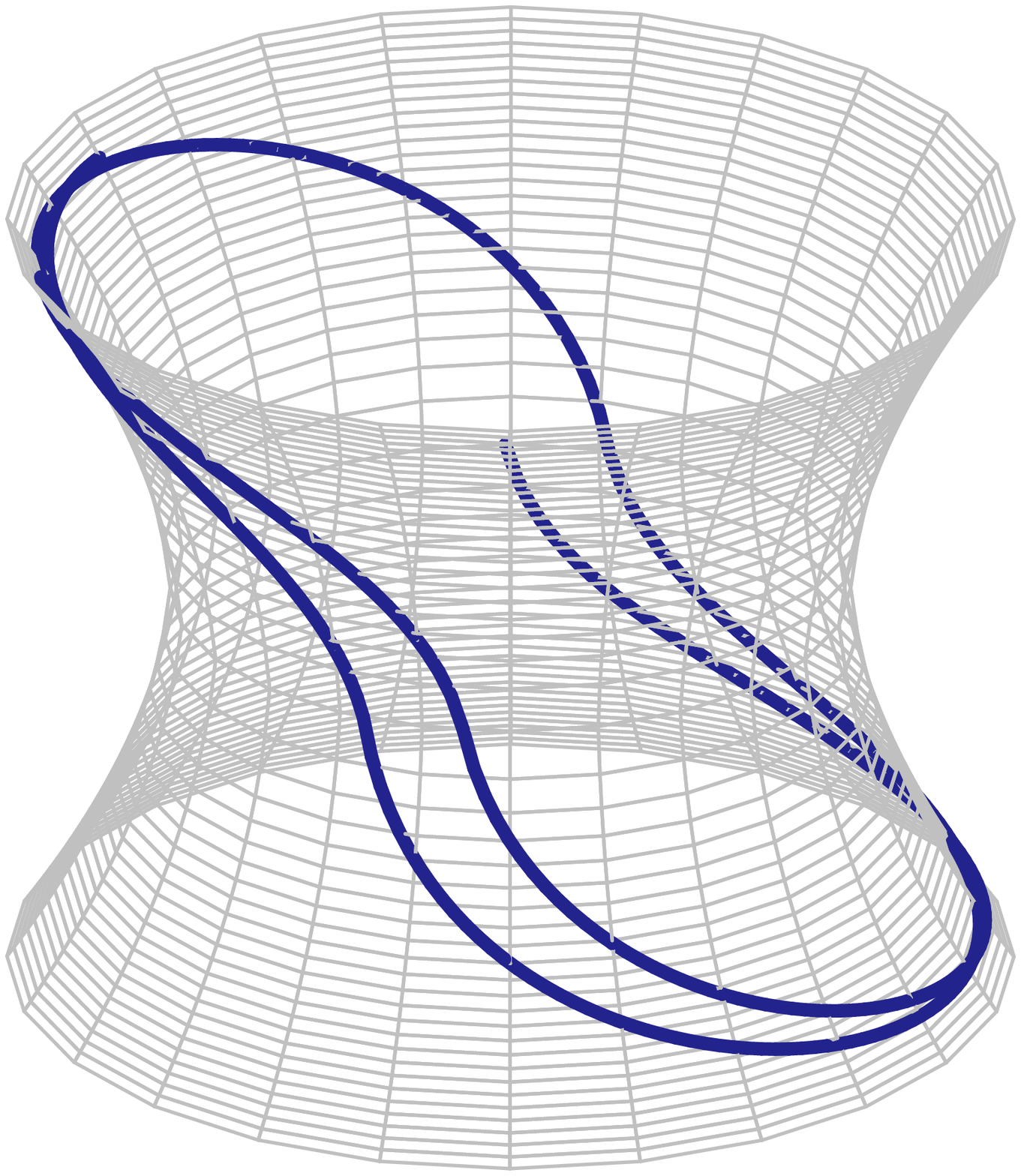}}
\subfigure[][$L^2=4.5, E^2=0.949$. BO.]{\label{schw_orb2}\includegraphics[width=4.5cm]{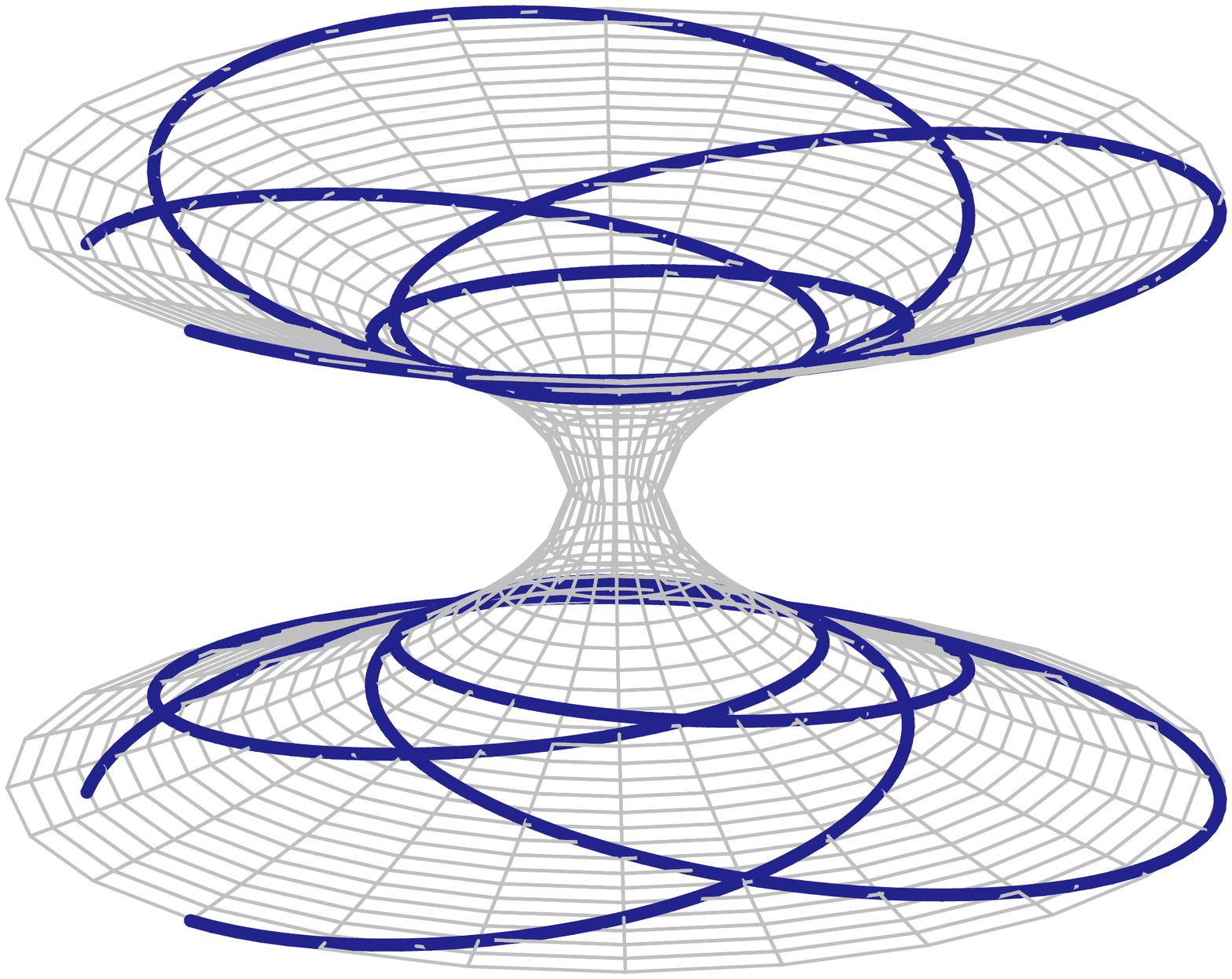}}
\subfigure[][$L^2=7.3, E^2=1.451867$. TWB.]{\label{schw_orb3}\includegraphics[width=3.2cm]{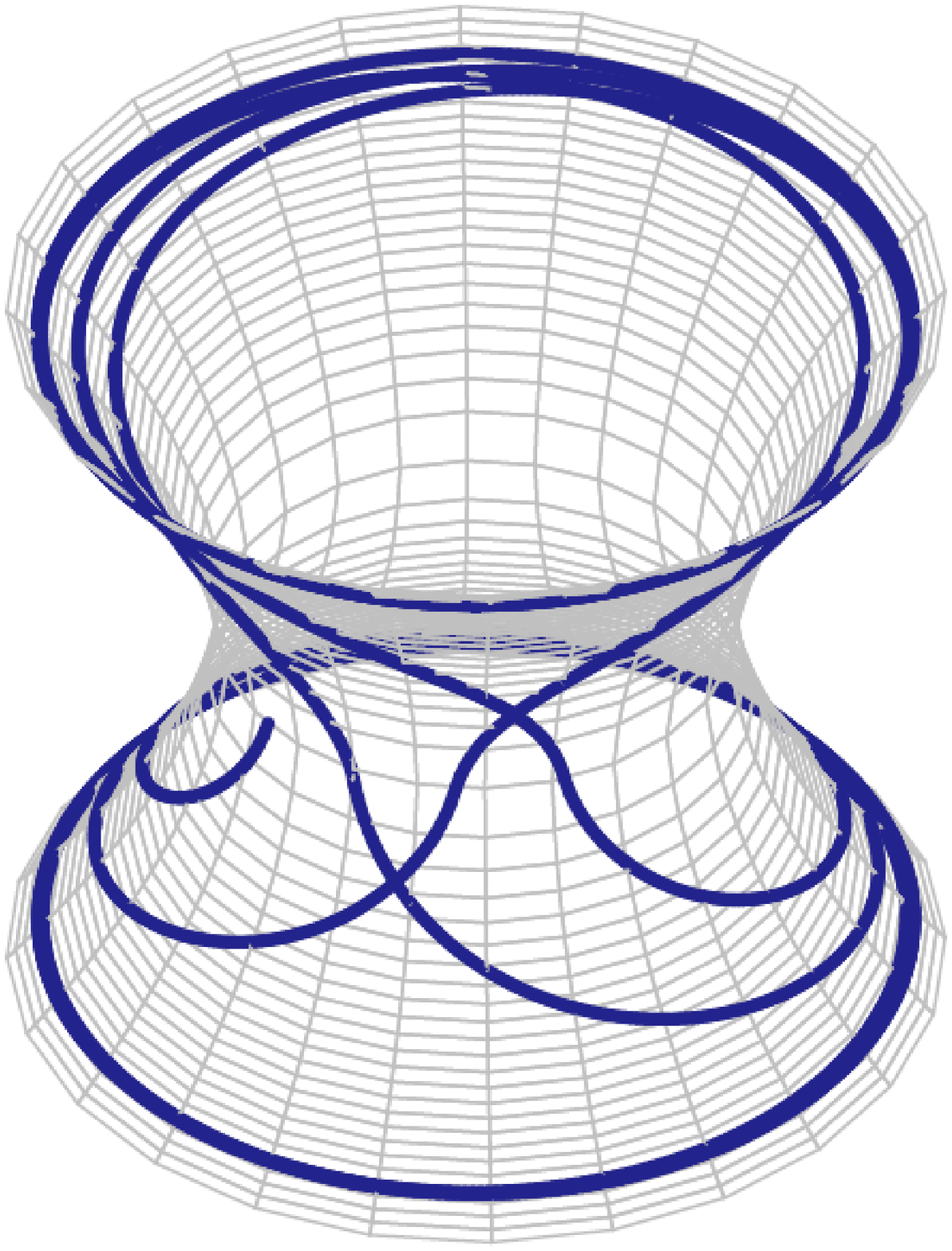}} \\
\subfigure[][$L^2=7.3, E^2=1.451867$. EO.]{\label{schw_orb4}\includegraphics[width=4.5cm]{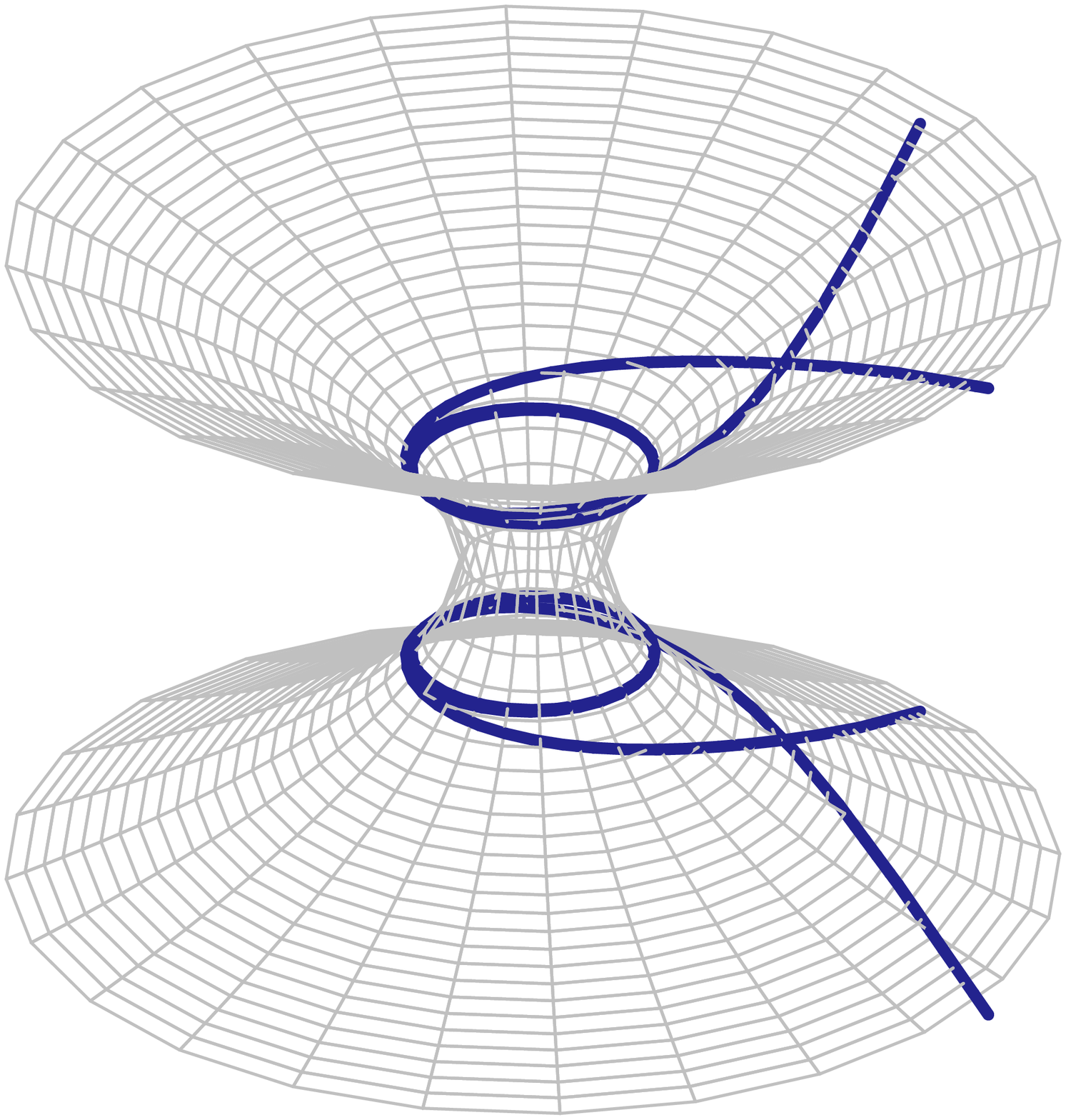}}
\subfigure[][$L^2=4.3, E^2=1.04$. TWE.]{\label{schw_orb5}\includegraphics[width=4.5cm]{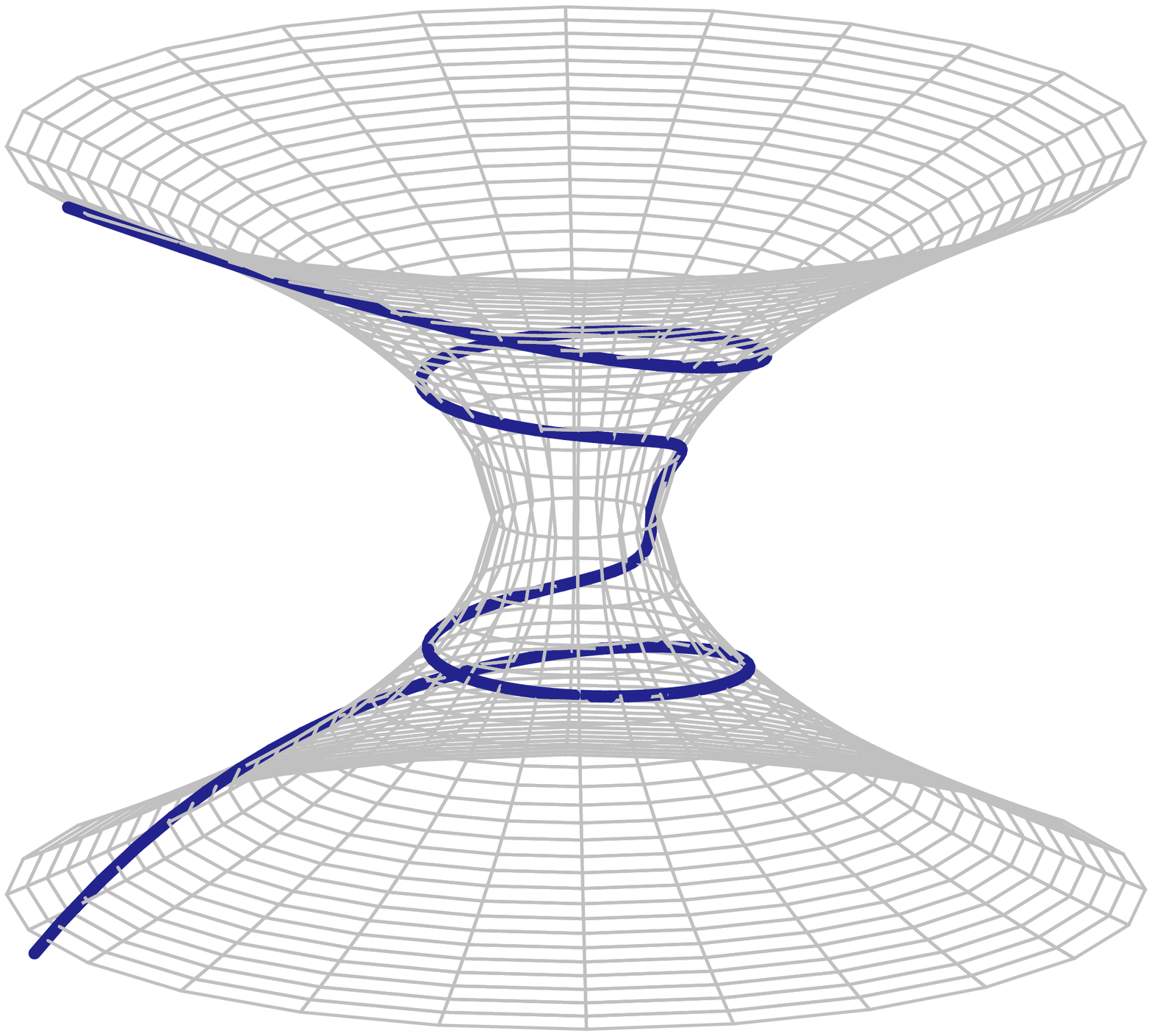}}
\subfigure[][$L^2=9, E^2=3$. TWE.]{\label{schw_orb6}\includegraphics[width=4.5cm]{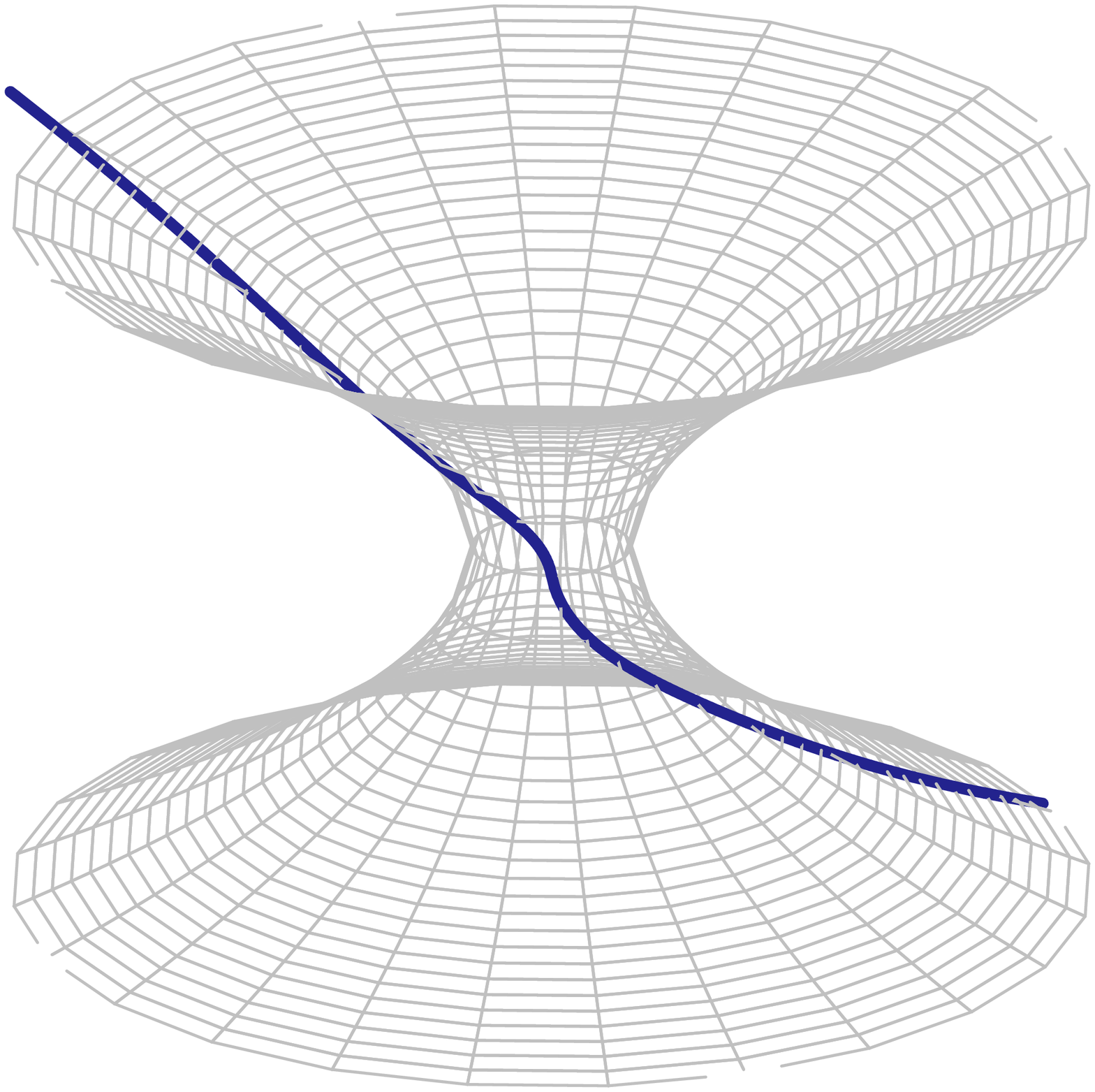}}
\end{center}
\caption{Orbits of a test particle with $\delta=1$ for different values of angular momentum and energy in the thin-shell Schwarzschild wormhole spacetime. The two-world-bound (TWB) and two-world-escape (TWE) orbits extend from the upper to the lower part of the Universe connected by a wormhole with throat parameter $b_0=1.01$. \label{fig:schworb}}
\end{figure*}

Geodesics of massive test particles described by the solution~\eqref{eq:schw6} are shown in Fig.\ref{fig:schworb}. In this Figure, the two-world bound TWB orbit~\subref{schw_orb1} and the bound orbit BO~\subref{schw_orb2} correspond to  type $D$. The two-world bound TWB orbit~\subref{schw_orb3} and the escape orbit EO~\subref{schw_orb4} are of type $C$, and the two-world-escape orbits TWE~\subref{schw_orb5} and~\subref{schw_orb6} are of type $A$. The orbits are embedded into the three-dimensional space with the coordinates $(x,y,z)$, where $x=r\cos(\varphi)$, $y=r\sin(\varphi)$, and $r$ and $z$ are given by~\eqref{eq:schw6} and~\eqref{schw:emb} correspondingly. 

\begin{figure*}[th!]
\begin{center}
\subfigure[][$L^2=4.5, E^2=0.6665$. TWB.]{\label{schw_orb1_light}\includegraphics[width=4.5cm]{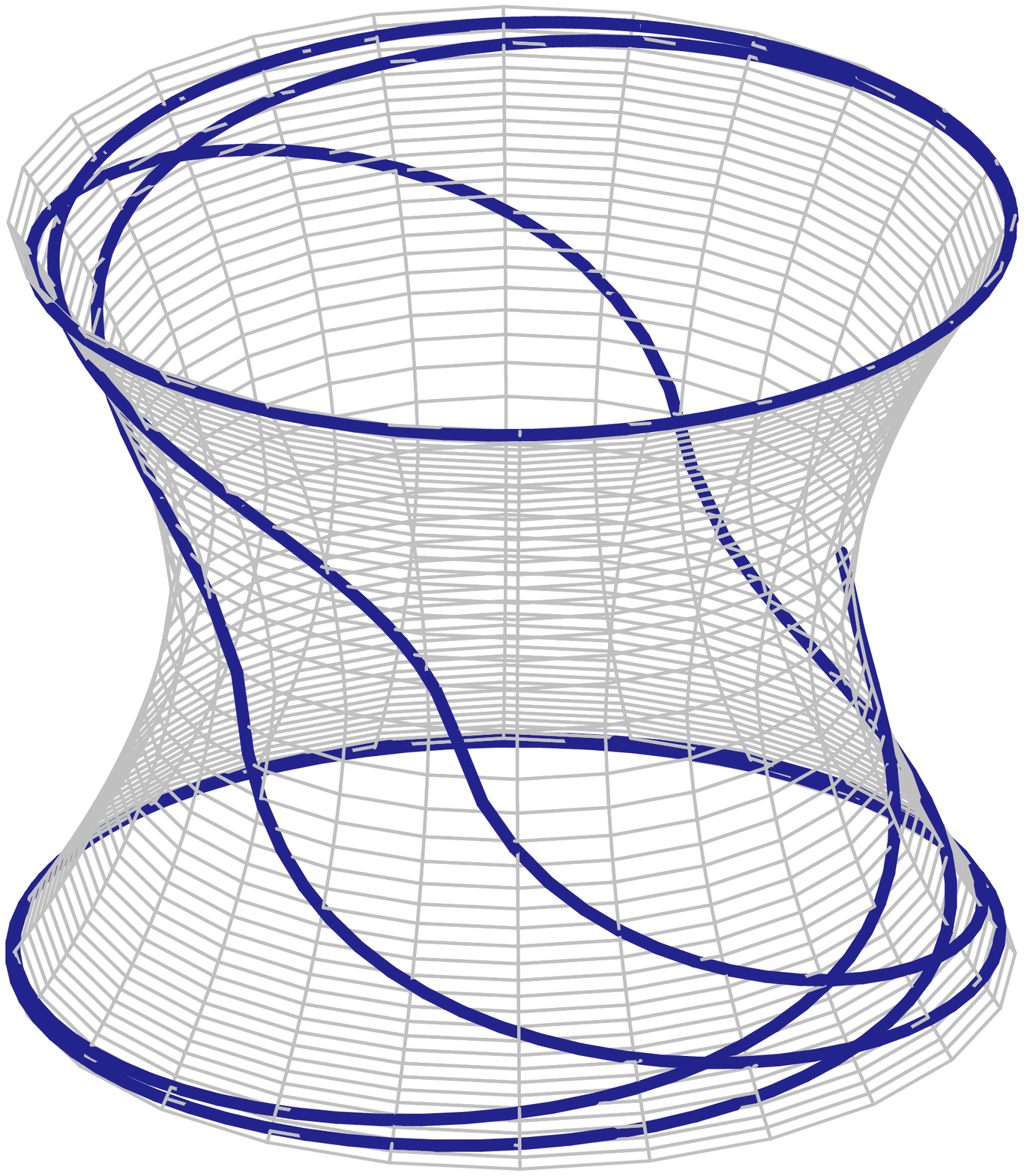}}
\subfigure[][$L^2=4.5, E^2=0.6665$. EO.]{\label{schw_orb2_light}\includegraphics[width=4.5cm]{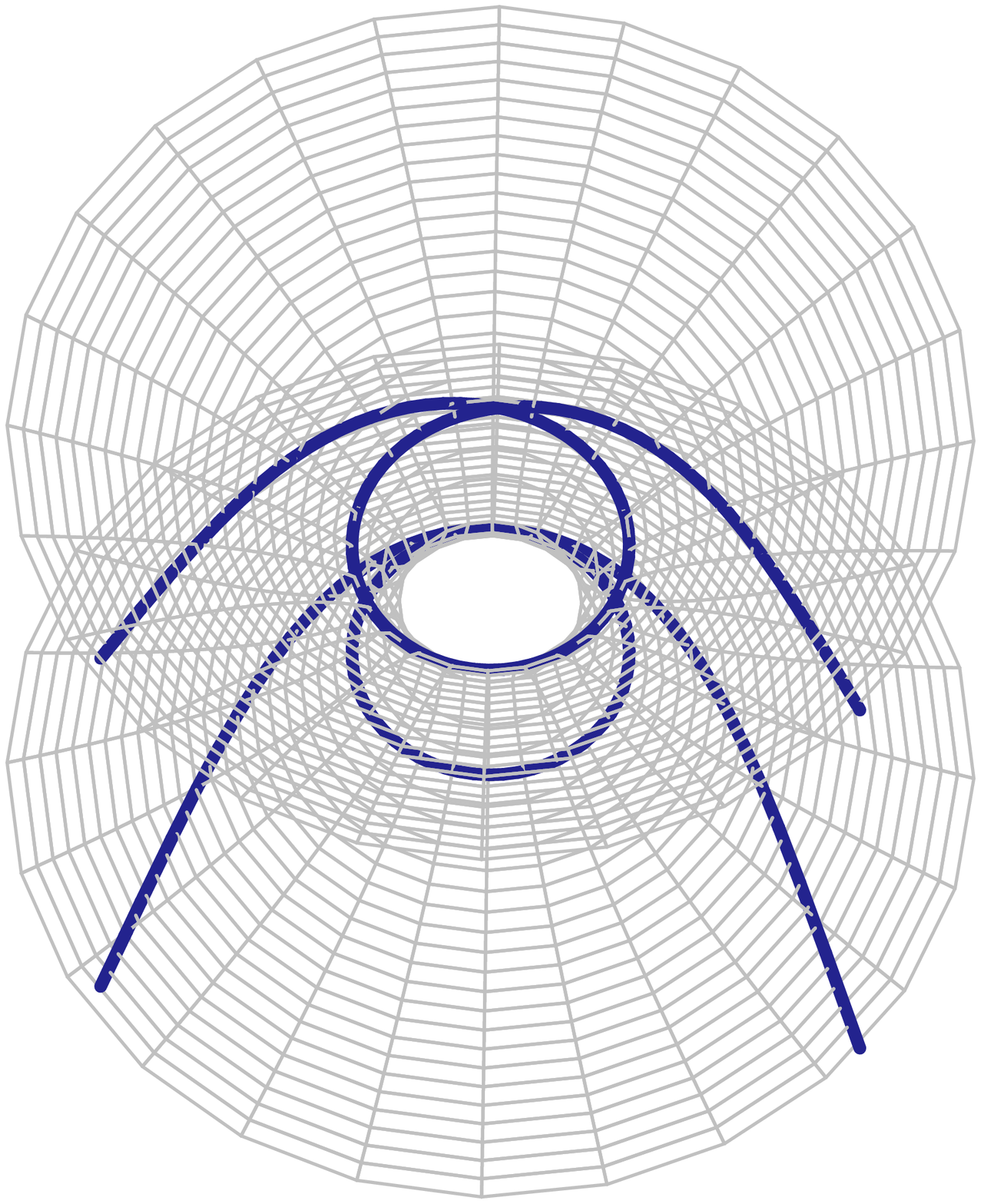}}
\subfigure[][$L^2=4.5, E^2=0.6667$. TWE.]{\label{schw_orb3_light}\includegraphics[width=4.5cm]{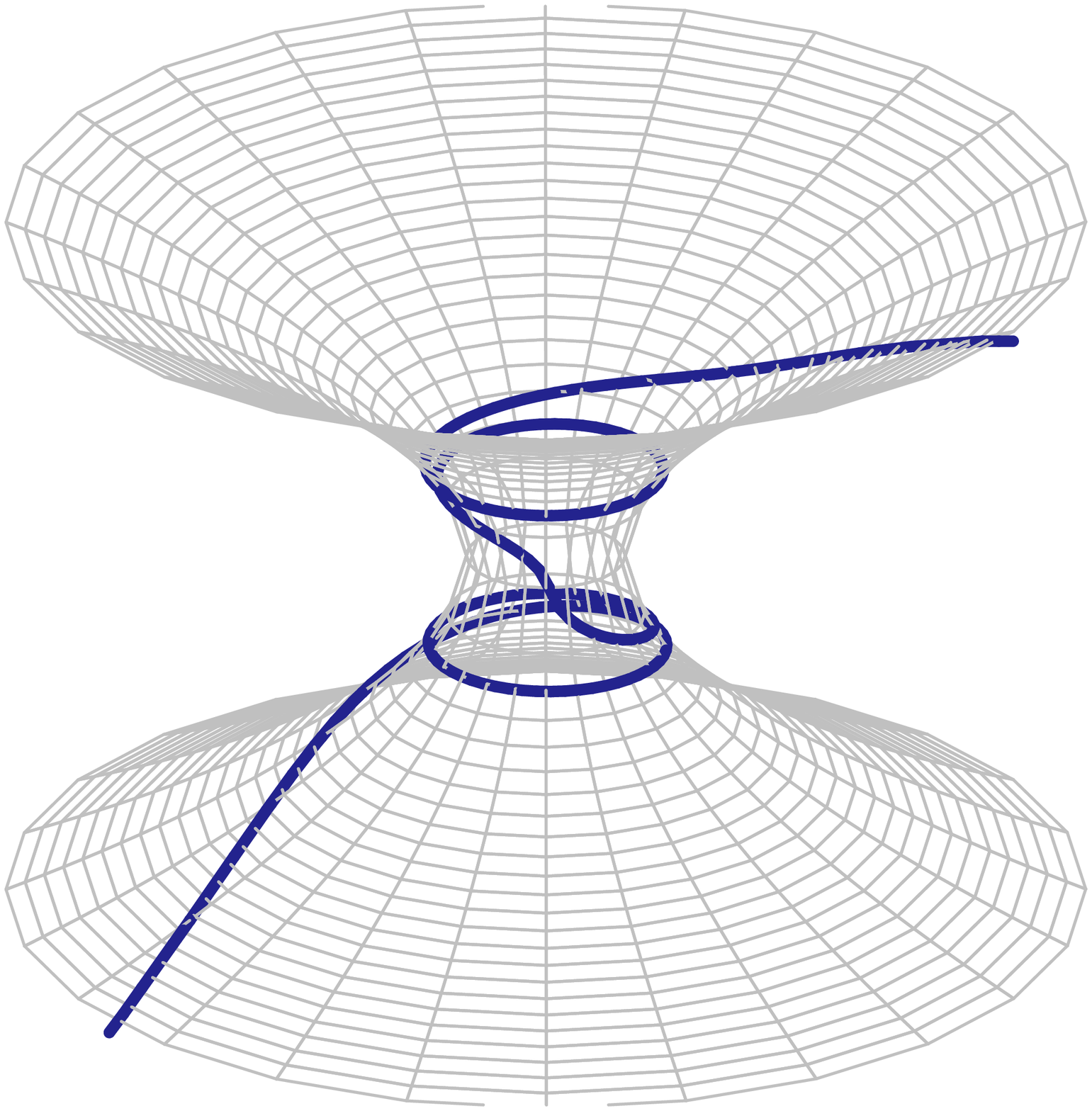}} \\
\end{center}
\caption{Orbits in the thin-shell Schwarzschild wormhole spacetime for a massless test particle ($\delta=0$). The two-world-bound (TWB) and two-world-escape (TWE) orbits are located in the upper and lower parts of the Universe which are connected by a wormhole with throat parameter $b_0=1.01$. \label{fig:schworb_light}}
\end{figure*}

\paragraph{Geodesics for $\delta=0$.}
Similar to the Schwarzschild black hole spacetime no planetary bound orbits exist in the Schwarzschild thin-shell wormhole spacetime. The orbits of type $A$ (two-world escape TWE), $B$ (two-world bound TWB) and $C$ (two-world bound TWB and escape EO) are possible. We show some examples in Fig.\ref{fig:schworb_light}.

\section{Kerr wormhole}\label{sec:kerr}

\subsection{The geodesic equations}

We start with the Kerr metric in Boyer–Lindquist coordinates~\cite{Chandrasekhar83}
\begin{eqnarray}
ds^2 & = & - \left( 1- \frac{r}{\rho^2} \right) dt^2 + \frac{\rho^2}{\Delta} dr^2 + \rho^2 d\vartheta^2 - \frac{2ar}{\rho^2}\sin^2\vartheta d\varphi dt \nonumber \\ 
     & + & \left( r^2+a^2+ \frac{a^2r}{\rho^2}\sin^2\vartheta \right)\sin^2\vartheta d\varphi^2   \ ,
\end{eqnarray}
where $\Delta=r^2+a^2-r$ and $\rho^2=r^2+a^2\cos^2\vartheta$. The radial coordinate $r$ and the rotation parameter $a$ are normalized to the mass parameter $2M$ and are dimensionless.

The Lagrangian for a free test particle following a geodesic in the equatorial plane $\vartheta=\frac{\pi}{2}$ yields
\begin{equation}
2\mathcal{L} = - \left( 1- \frac{1}{r} \right) \dot{t}^2 + \frac{r^2}{\Delta} \dot{r}^2 - \frac{2a}{r} \dot{\varphi} \dot{t} + \left( r^2+a^2+ \frac{a^2}{r} \right) \dot{\varphi}^2   \,\,\,\, {\rm with} \,\,\,\, 2 \mathcal{L} = - \delta \label{eq:kerrLagrangian} \ ,
\end{equation}
where $\delta=0$ for null geodesics and $\delta=1$ for time-like geodesics, and $\dot{x^\alpha}$ is the derivative with respect to the affine parameter $\lambda$.

The conserved and dimensionless energy $E = - \frac{\partial \mathcal{L} }{\partial \dot{t}}$ and angular momentum $L = \frac{\partial \mathcal{L}}{\partial \dot{\varphi}}$ of a free test particle are
\begin{eqnarray}
&& E = \left( 1- \frac{1}{r} \right) \dot{t} + \frac{a}{r} \dot{\varphi} \label{eq:kerrE} \ , \\
&& L = - \frac{a}{r} \dot{t} + \left( r^2+a^2+ \frac{a^2}{r} \right) \dot{\varphi} \label{eq:kerrL}  \ .
\end{eqnarray}

Solving the equations~\eqref{eq:kerrE} and~\eqref{eq:kerrL} with respect to $\dot{t}$ and $\dot{\varphi}$ and substituting them into the Lagrangian~\eqref{eq:kerrLagrangian} we obtain the differential equations describing the motion of a free test particle in the equatorial plane~\cite{Chandrasekhar83,Oneil}
\begin{eqnarray}
     & \left(\frac{dr}{d\gamma}\right)^2 & = E^2r^4 + r^2(a^2E^2-L^2) - \delta\Delta r^2+ r(aE-L)^2  \ , \,\,\, \label{eq:kerr-r} \\
     & \frac{d\varphi}{d\gamma}  & = \frac{r}{\Delta}\left( L r + (aE - L) \right)  \ . \label{eq:kerr-phi}
\end{eqnarray}
Here $\gamma$ is a new affine parameter related to $\lambda$ by $d\gamma = r^{-2} d\lambda $.

The ergosphere, defined by the condition $g_{tt}=0$, is located at $r=1$ for the equatorial plane. The condition $g_{rr}\rightarrow \infty$ corresponding to the vanishing of $\Delta$ defines the two horizons of the Kerr spacetime: $h_{1,2}=(1\mp\sqrt{1-4a^2})/2$.

In the following we will solve the diffential equations~\eqref{eq:kerrE} and~\eqref{eq:kerrL} using the theory of elliptic functions~\cite{Markush}.

\paragraph{Radial equation.}

By the transformation $r=a_3 (4y-a_2/3)^{-1}$, the equation~\eqref{eq:kerr-r} reduces to the Weierstrass form~\cite{Markush}
\begin{equation}
\left( \frac{dy}{d\gamma} \right)^2 = 4 y^3 - \nu_2 y - \nu_3 = P_3(y) \ , \label{eq:kerr-r2} 
\end{equation}
where $\nu_2=(a_2^2/3-a_1a_3)/4$ and $\nu_3=a_1a_2a_3/48-a_0a_3^2/16-a_2^3/216$. The coefficients $a_i$ read: 
\begin{eqnarray}
 && a_3=(aE-L)^2 ,  \,\, a_2=a^2(E^2-\delta)-L^2 \ , \nonumber \\  
 && a_1=\delta ,  \,\,\,  a_0=E^2-\delta \ . 
\end{eqnarray}

The solution of the differential equation~\eqref{eq:kerr-r2} is the Weierstrass $\wp$-function $y=\wp(\gamma-\gamma^\prime)$, where $\gamma^\prime=\gamma_0+\int^\infty_{y_0}(4y^3-\nu_2y-\nu_3)^{-1/2}dy$ is a constant defined by the initial conditions of the geodesic. With this result, the final solution for the coordinate $r$ yields~\cite{Chandrasekhar83,Oneil,Markush}:
\begin{equation}
r = \frac{3 a_3}{12 \wp(\gamma-\gamma^\prime) - a_2} \ . \label{eq:kerr-r3} 
\end{equation}

\paragraph{Azimuthal equation.}

The transformation $r=a_3 (4y-a_2/3)^{-1}$, reduces the equation~\eqref{eq:kerr-phi} to a sum of two differentials of the third kind:
\begin{equation}
   d\varphi = \sum^2_{i=1} \frac{K_i}{y-y_i} \frac{dy}{\sqrt{P_3(y)}} \label{eq:kerr-phi2} \ ,
\end{equation}
where the constants $K_1$ and $K_2$ are defined as
\begin{equation}
   K_i = \frac{(-1)^{i-1}}{\sqrt{1-4a^2}} \frac{a_3}{4}  \left( L + \frac{(aE-L)(2a^2)^{-1}}{\left(1+(-1)^{i-1}\sqrt{1-4a^2}\right)^{-1}} \right) \label{eq:k1k2} \ ,
\end{equation}
and $i=1,2$.

To proceed, we introduce $w=\gamma-\gamma^\prime$ and replace $y$ in~\eqref{eq:kerr-phi2} by $y=\wp(w)$. This yields
\begin{equation}
   d\varphi= \sum^2_{i=1} \frac{K_i}{\wp(w) - \wp_{{y_i}}} dw \label{eq:kerr-phi3} \ ,
\end{equation}
where $\wp_{{y_i}}$, $i=1,2$, is defined by the equation $\wp_{{y_i}}=\wp(w_{y_i})=y_i$.

The integration of~\eqref{eq:kerr-phi3} in terms of elliptic $\zeta$ and $\sigma$ functions reads~\cite{Markush,GK10,MPaeqb,KKHL10}

\begin{equation}
\varphi = \varphi_0 + 
\sum^2_{i=1}\frac{K_i}{\wp^\prime(w_{y_i})}
\Biggl( 2\zeta(w_{y_i})(w-w_{0}) + \ln\frac{\sigma(w-w_{y_i})}{\sigma(w+w_{y_i})}
- \ln\frac{\sigma(w_{0}-w_{y_i})}{\sigma(w_{0}+w_{y_i})} \Biggr)  \ , \label{eq:kerr-phi4}  \ 
\end{equation}
where $w_0=w(\gamma_0)$.

\subsection{Construction of the wormhole and embedding}\label{sec:kerr_emb}

The throat of the Kerr wormhole lies behind the horizons: $b_0>h_2$. Following~\cite{KashSush2011} we assume that the thin-shell Kerr-wormhole with the throat parameter $b_0$ does not depend on the proper time and, thus, is constant. Similar to the Sec.~\ref{sec:schw_emb} we use the surgery method to construct the thin-shell Kerr-wormhole. Namely, we cut the Kerr spacetime at some $b_0>h_2$ and glue two copies of the manifold $M_{1,2}=(t,r,\vartheta,\varphi | r\geq b_0)$ onto each other. Identifying the boundaries of these copies we get a new geodesically complete manifold with two regions connected by a wormhole with the throat characterized by the parameter $b_0$. The new coordinate $|l|=|r-b_0|$ running from $-\infty$ to $\infty$ and vanishing at the throat is a practical coordinate to portray the effective potential~\eqref{eq:VeffKerr} (see Fig.\ref{fig:kerrpot} and Fig.\ref{fig:kerrpot2} in the Sec.\ref{sec:geod_kerr}).

To understand the Kerr-wormhole topology for the equatorial plane motion described by the functions~\eqref{eq:kerr-r3} and~\eqref{eq:kerr-phi4}, consider the two dimensional hypersurface $(t=const, \vartheta=\pi/2)$ described by the geometry: 
\begin{eqnarray}
ds_0 &=& \frac{r^2}{\Delta}dr^2 + \left( r^2 + a^2 + \frac{a^2}{r} \right)d\varphi^2 \nonumber \\ 
     &=& \frac{r^2}{\Delta}dr^2 + R^2 d\varphi^2 \label{kerr_emb1} \ ,
\end{eqnarray}
where $R^2=r^2 + a^2 + \frac{a^2}{r}$ and $\Delta=r^2+a^2-r$.
This two-dimensional hypersurface can be embedded into the Euclidean space given by
\begin{eqnarray}
ds_E &=& dR^2 + R^2 d\varphi^2 + dz^2 \nonumber \\
     &=& \left( \left(\frac{dR}{dr}\right)^2 + \left(\frac{dz}{dr}\right)^2 \right)dr^2 + R^2 d\varphi^2 \label{kerr_emb2} \,
\end{eqnarray}
in the cylindrical coordinates $(r,\varphi,z)$. 

Comparing the coefficients of $dr^2$ in~\eqref{kerr_emb1} and~\eqref{kerr_emb2} we find that the shape of the embedding diagram in the Euclidean space is given by the integral
\begin{equation}
z(r) = \int^r_{b_0} \sqrt{ \frac{{r^\prime}^2}{\Delta} - \frac{(2{r^\prime}^3-a^2)^2}{4{r^\prime}^3({r^\prime}^3+a^2({r^\prime}+1))} } d{r^\prime} \label{kerr_emb3} \,
\end{equation}
and is visualized in Fig.\ref{fig:kerrorb} in coordinates $(x,y,z)$, where $x=r\cos(\varphi)$ and $y=r\sin(\varphi)$ (to the discussion of the orbits we will come in the next section). The integral in~\eqref{kerr_emb3} is calculated numerically. For the embedding diagram we choose the initial value of $r$ to be $b_0$ and $\varphi\in[0,2\pi]$.

\subsection{Geodesics}\label{sec:geod_kerr}

We introduce the effective potential via equation~\eqref{eq:kerr-r}
\begin{equation}
\left( \frac{dr}{d\gamma} \right)^2 = r \left(r^3 + a^2 (r + 1) \right) \left( E - V^+_{\rm{eff}} \right)\left( E - V^-_{\rm{eff}} \right) \label{eq:Veff_in_Kerr} \ .
\end{equation}
Thus, the effective potential in the Kerr spacetime is given by~\cite{Chandrasekhar83,Oneil,Pugliese}
\begin{equation}
V_{\rm{eff}} \equiv V^{\pm}_{\rm{eff}}  = \frac{ aL\pm\sqrt{ r\Delta\left( rL^2 + \delta\left(r^3 + a^2(r+1)\right)  \right)} }{ r^3 + a^2(r+1) } \label{eq:VeffKerr} \ . 
\end{equation}
The intersections of the function $E$ and $V_{\rm{eff}}$ specify the turning points of the motion. 

\begin{figure}[th!]
\subfigure[][$b_0=0.96$]{\label{kerr_pot1-posneg}\includegraphics[width=7.5cm]{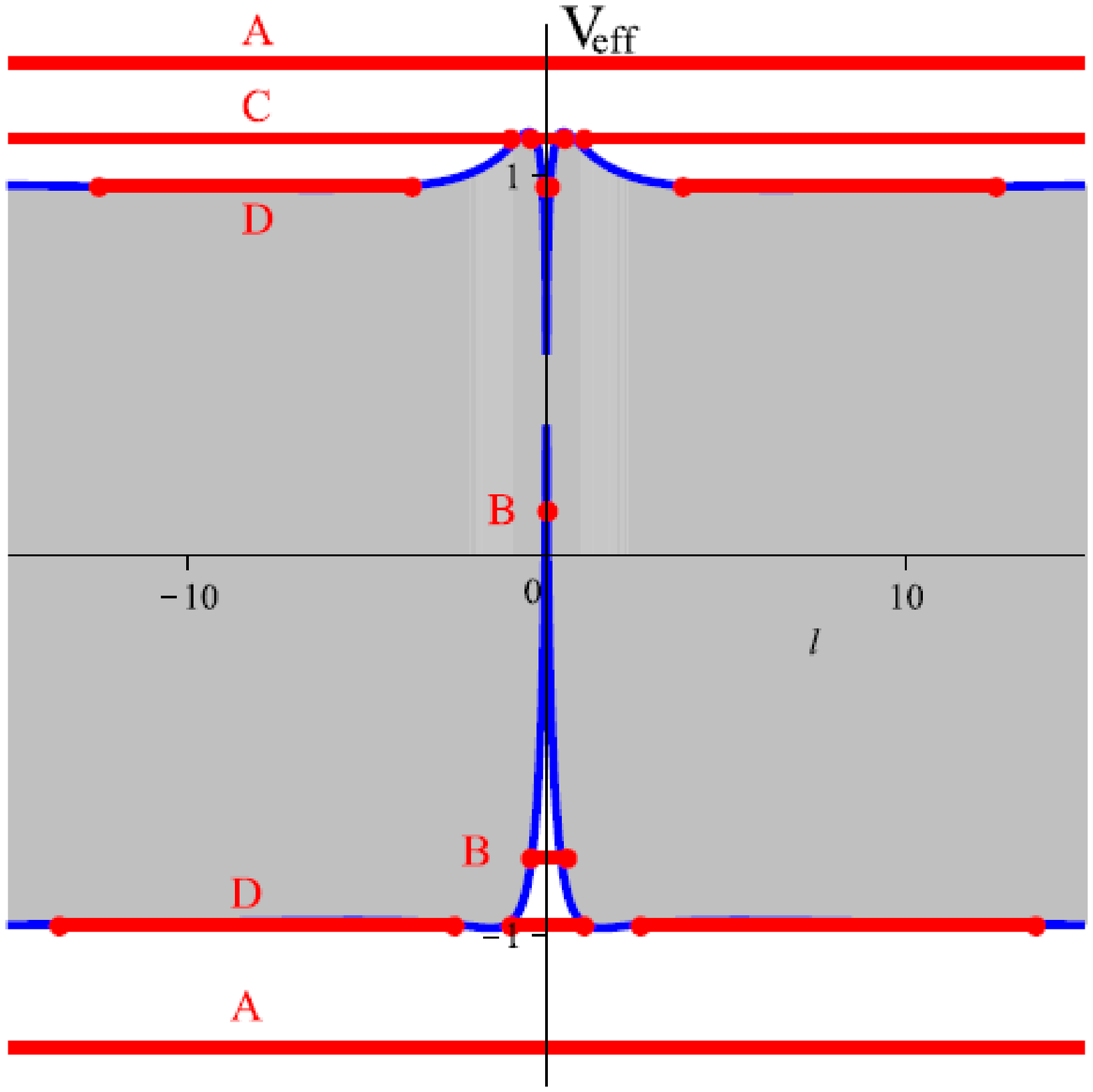}} 
\subfigure[][Detailed representation of the potential~\subref{kerr_pot1-posneg}.]{\label{kerr_pot1}\includegraphics[width=8.5cm]{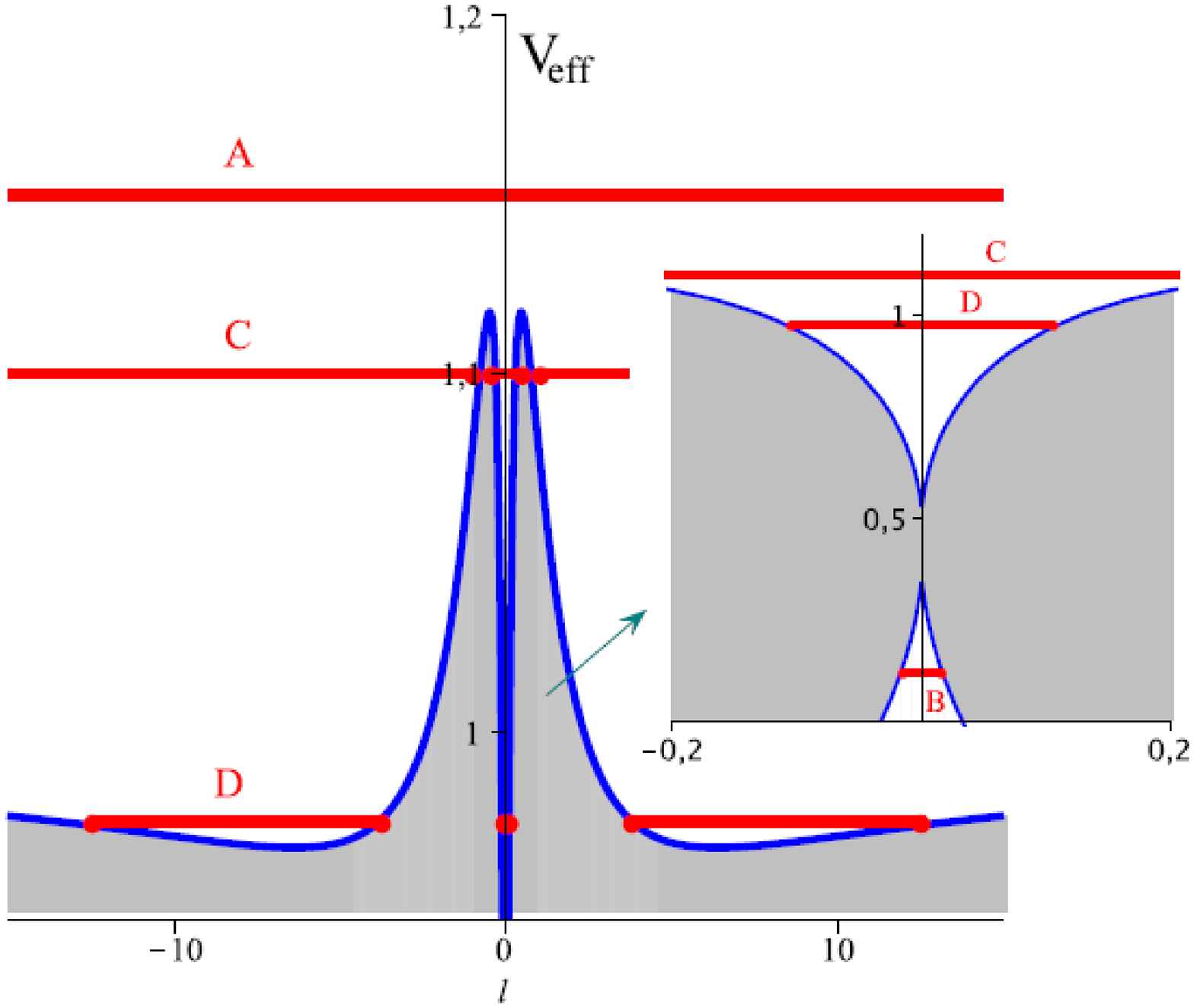}} \\
\subfigure[][$b_0=2.0$]{\label{kerr_pot2}\includegraphics[width=8.5cm]{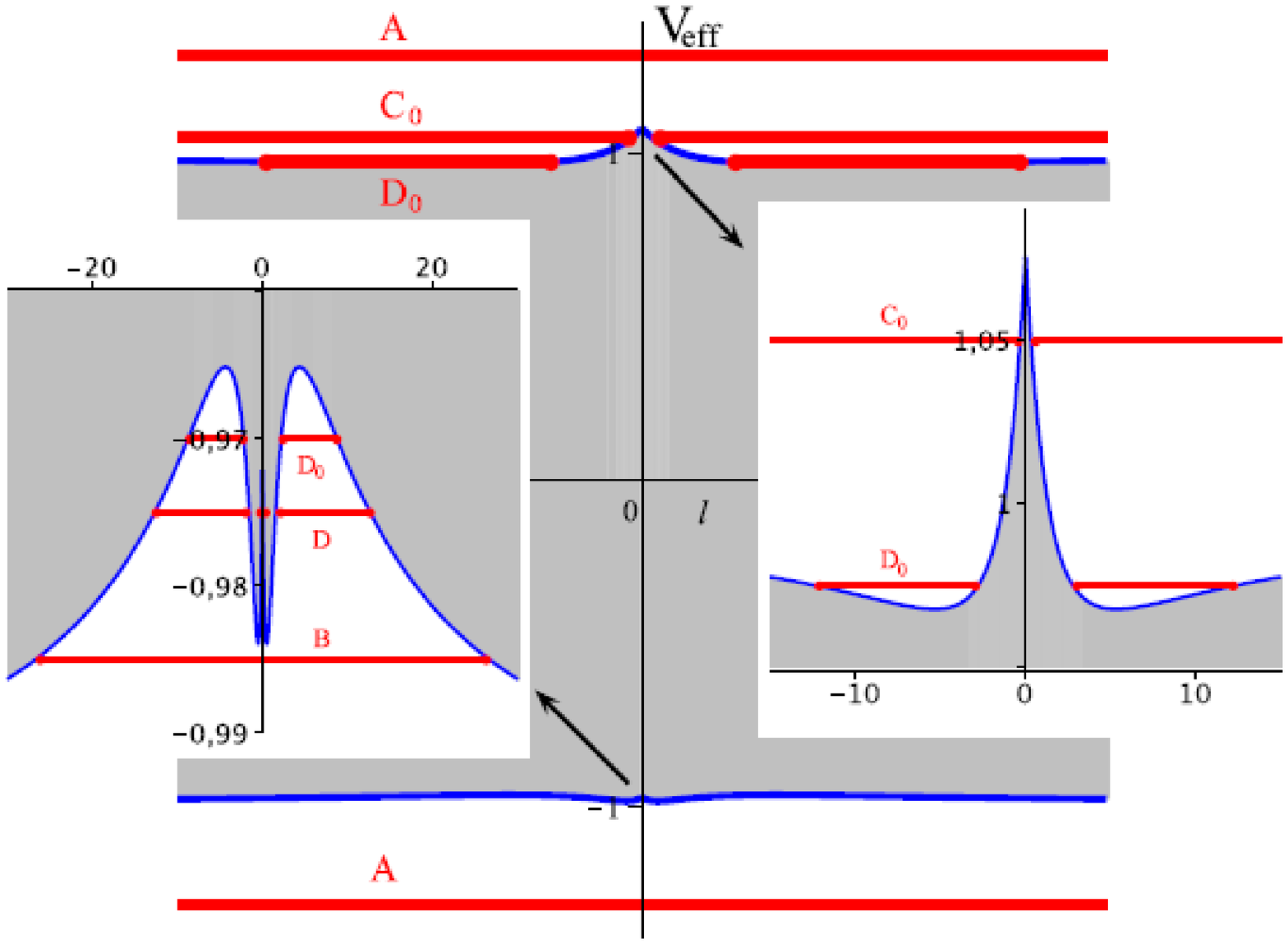}} 
\subfigure[][$b_0=10.0$]{\label{kerr_pot3}\includegraphics[width=8.5cm]{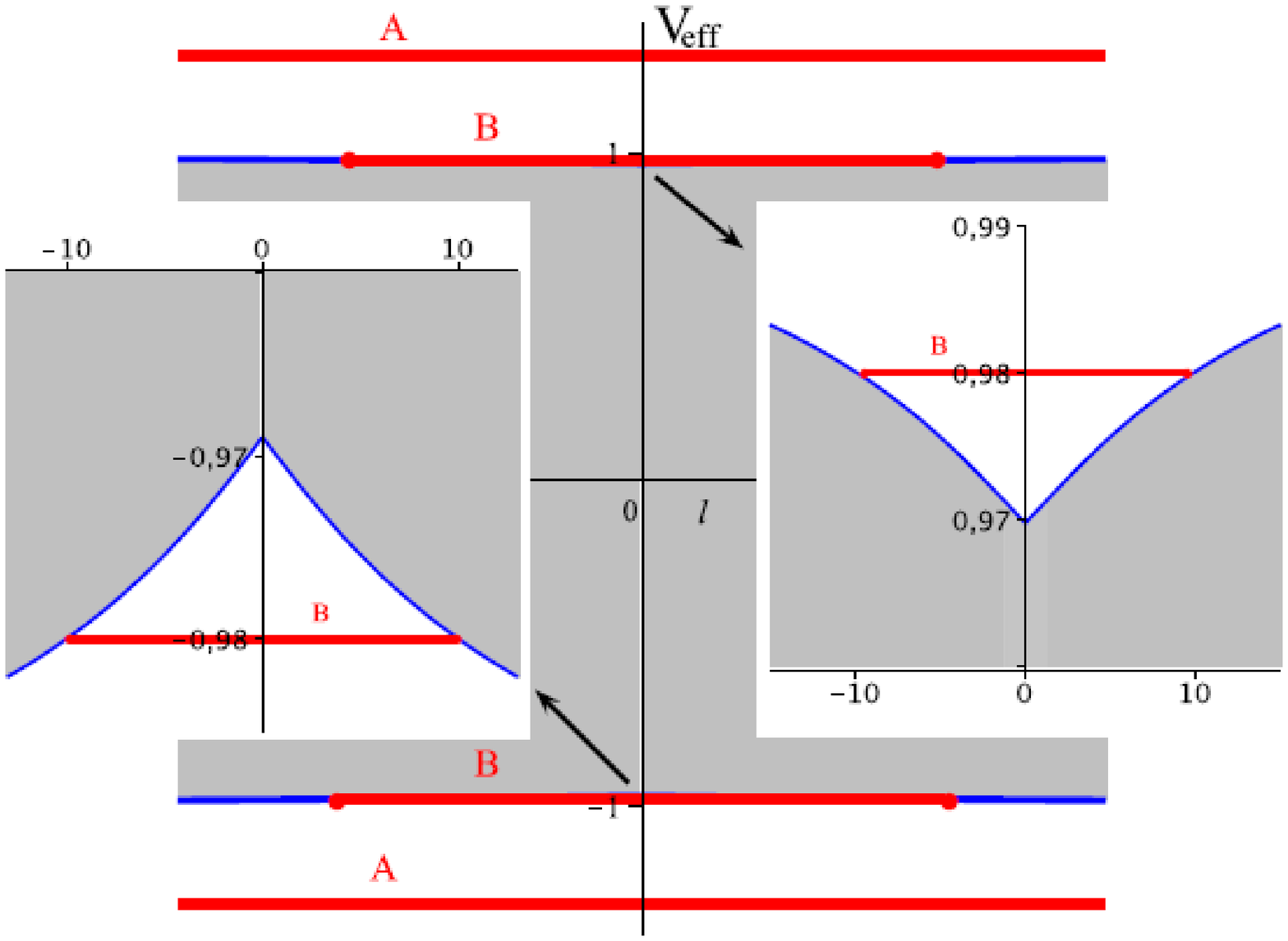}}
\caption{The effective potential (eq.~\eqref{eq:VeffKerr}) for $\delta=1$ and $L=2.1$. $a=0.2$ in the Kerr wormhole spacetime. $b_0>h_2$. Filled grey regions denote forbidden energy values. See Sec.\ref{sec:geod_kerr} for details. \label{fig:kerrpot}}
\end{figure}

The graphical representation of the effective potential~\eqref{eq:VeffKerr} is shown in Fig.\ref{fig:kerrpot}, where the coordinate $|l|=|r-b_0|$ is used. The regions which
do not satisfy the condition $E\geq V^+_{\rm{eff}}$ and $E\leq V^-_{\rm{eff}}$ are forbidden and colored grey.

\begin{figure*}[th!]
\subfigure[][$a=0.2$, $b_0=0.96$]{\label{kerr_pot4}\includegraphics[width=8.5cm]{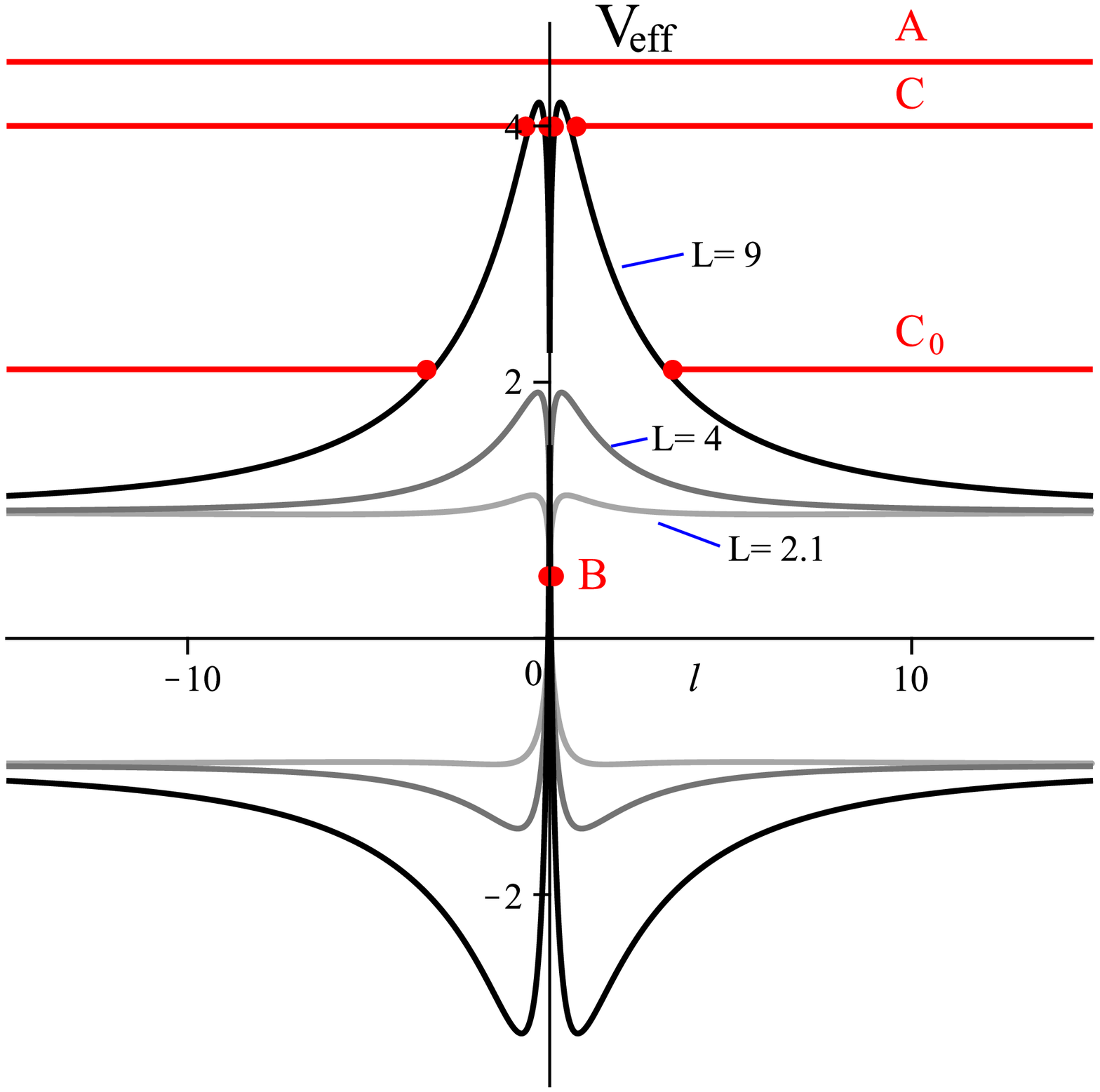}}
\subfigure[][$L=2.1$, $b_0=h_2+0.01$]{\label{kerr_pot5}\includegraphics[width=8.5cm]{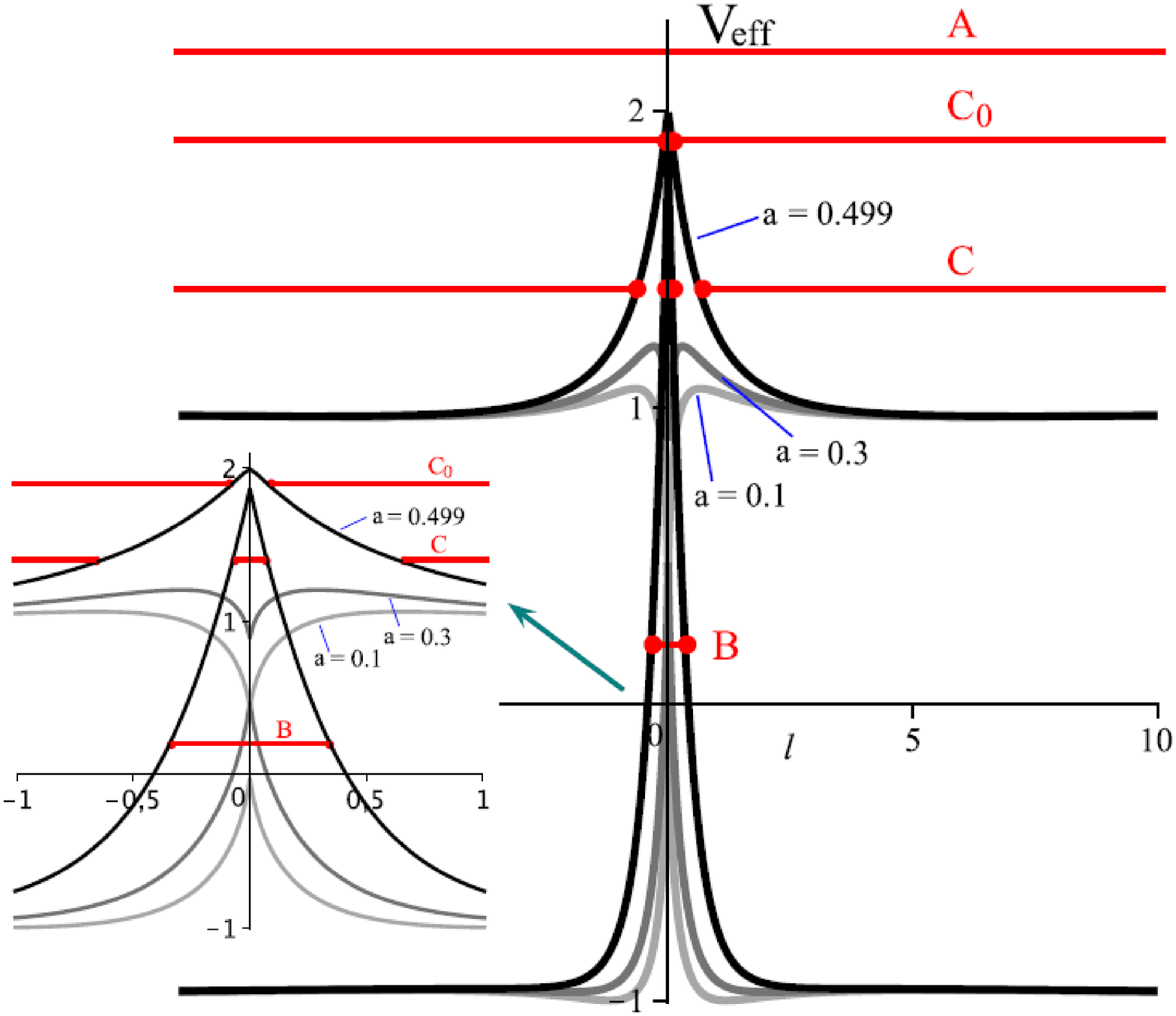}}
\caption{The effective potential (eq.~\eqref{eq:VeffKerr}) for $\delta=1$ and different values of $L$~\subref{kerr_pot4} and $a$~\subref{kerr_pot5} in the thin-shell Kerr wormhole spacetime. $b_0>h_2$. When keeping $a$ constant and increasing $L$  in~\subref{kerr_pot4}, we see that planetary bound orbits (i.e. of type $D$) do not exist (as expected from the Kerr black hole spacetime). When keeping $L$ constant and increasing $a$ in~\subref{kerr_pot5}, we observe the form of the effective potential change. We show the possible orbit types (for positive energies) for the potential in black for an almost critical rotation parameter $a=0.499$ and the value $b_0$ a bit larger than the horizon. In order not to overcharge the picture, we do not fill the forbidden regions with color (cp. Fig.\ref{fig:kerrpot}).  \label{fig:kerrpot2}}
\end{figure*}

The orbit types in the thin-shell Kerr-wormhole are very similar to those in the thin-shell Schwarzschild-wormhole. Namely,
\begin{itemize}
\item[Type $A$] two-world escape orbit TWE which connects two regions of the Universe. In the Kerr black hole spacetime this orbit would continue into the negative radial coordinate region.
\item[Type $B$] two-world bound orbits TWB. A test particle moves on a bound orbit which is partly located in the lower (respectively, upper) Universe, so that, a particle flies through the throat.
\item[Type $C$] here two-world bound TWB and escape orbits EO are possible. EOs exist in both parts of the Universe. For growing values of the throat parameter $b_0$ this type reduces to $C_0$ containing only EOs. See Fig.\ref{kerr_pot2}.
\item[Type $D$] two-world bound TWB and bound orbits BO are possible. BOs exist in both parts of the Universe. For growing values of the throat parameter $b_0$ the type $D$ reduces to $D_0$ containing only BOs (see Fig.\ref{kerr_pot2}).
\end{itemize}

For a further growing throat parameter $b_0$, TWB (type $B$) and TWE (type $A$) exist (see Fig.\ref{kerr_pot3}). For larger $b_0$ only the TWE (type $A$) orbits remain. In Fig.\ref{fig:kerrpot2} we make $L$ (in~\subref{kerr_pot4}) and $a$ (in~\subref{kerr_pot5}) change. In the first case, similar to the Kerr black hole spacetime, for large angular momenta of a test particle no planetary bound orbits exist. In the second case the form of the effective potential transforms when $a$ tends to the critical value $a_{\rm crit}=0.5$ when the horizons in the Kerr black-hole spacetime merge.

\begin{figure*}[th!]
\begin{center}
\subfigure[][$E=1.11722, L=2.1$. TWB.]{\label{kerr_orb1}\includegraphics[width=4.5cm]{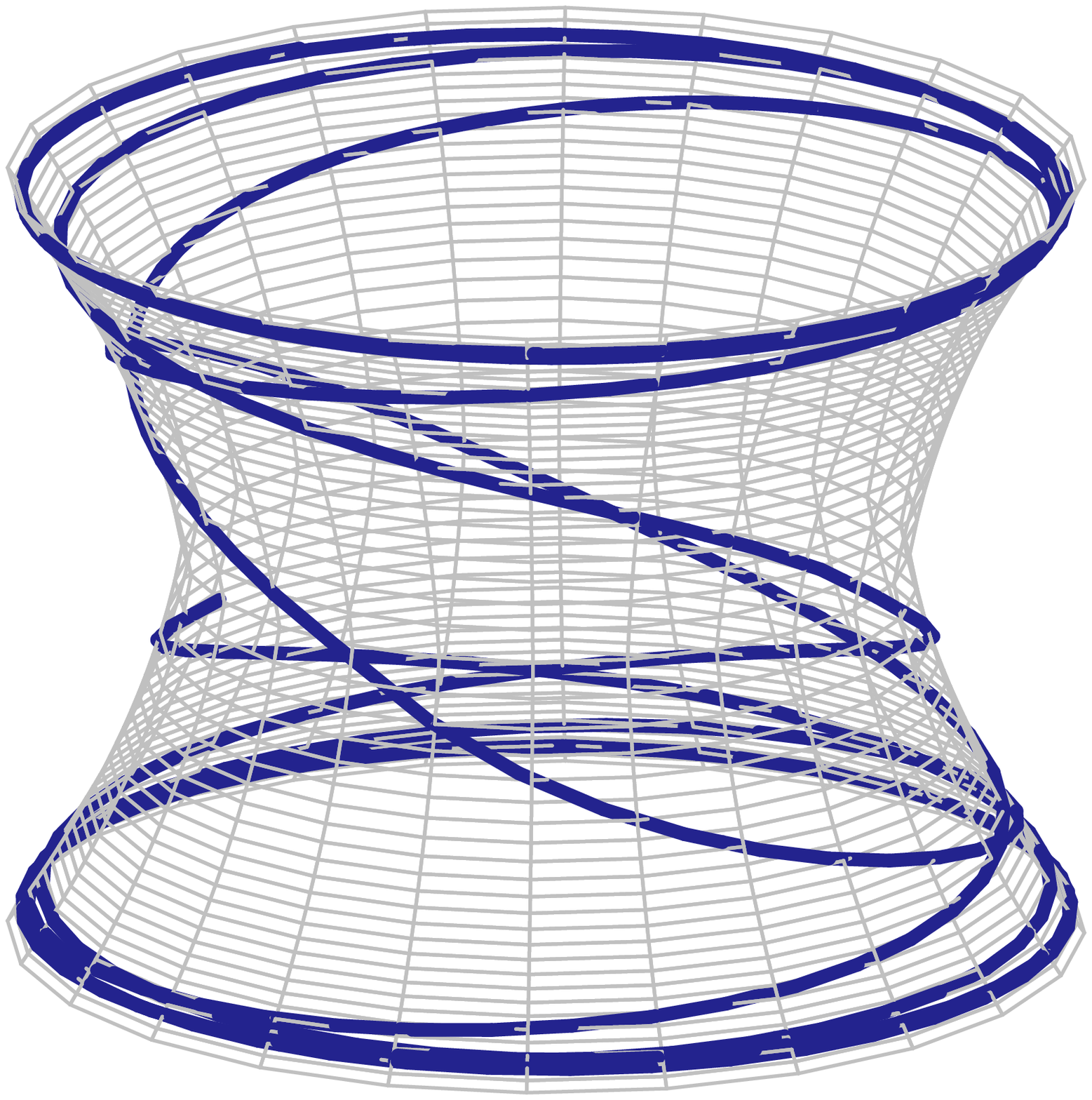}}
\subfigure[][$E=1.11722, L=2.1$. EO.]{\label{kerr_orb2}\includegraphics[width=4.5cm]{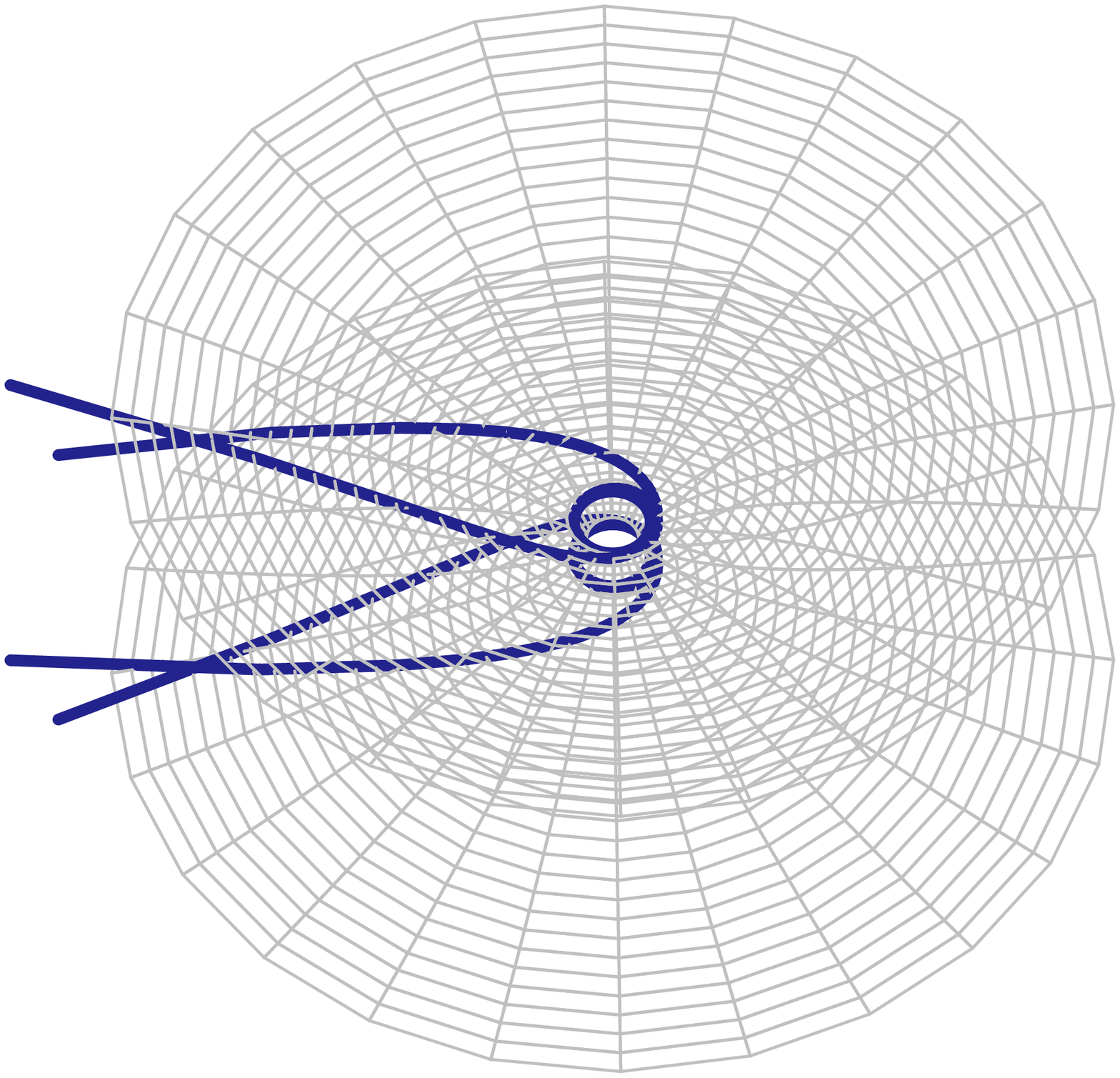}}
\subfigure[][$E=1.11723, L=2.1$. TWE.]{\label{kerr_orb3}\includegraphics[width=4.5cm]{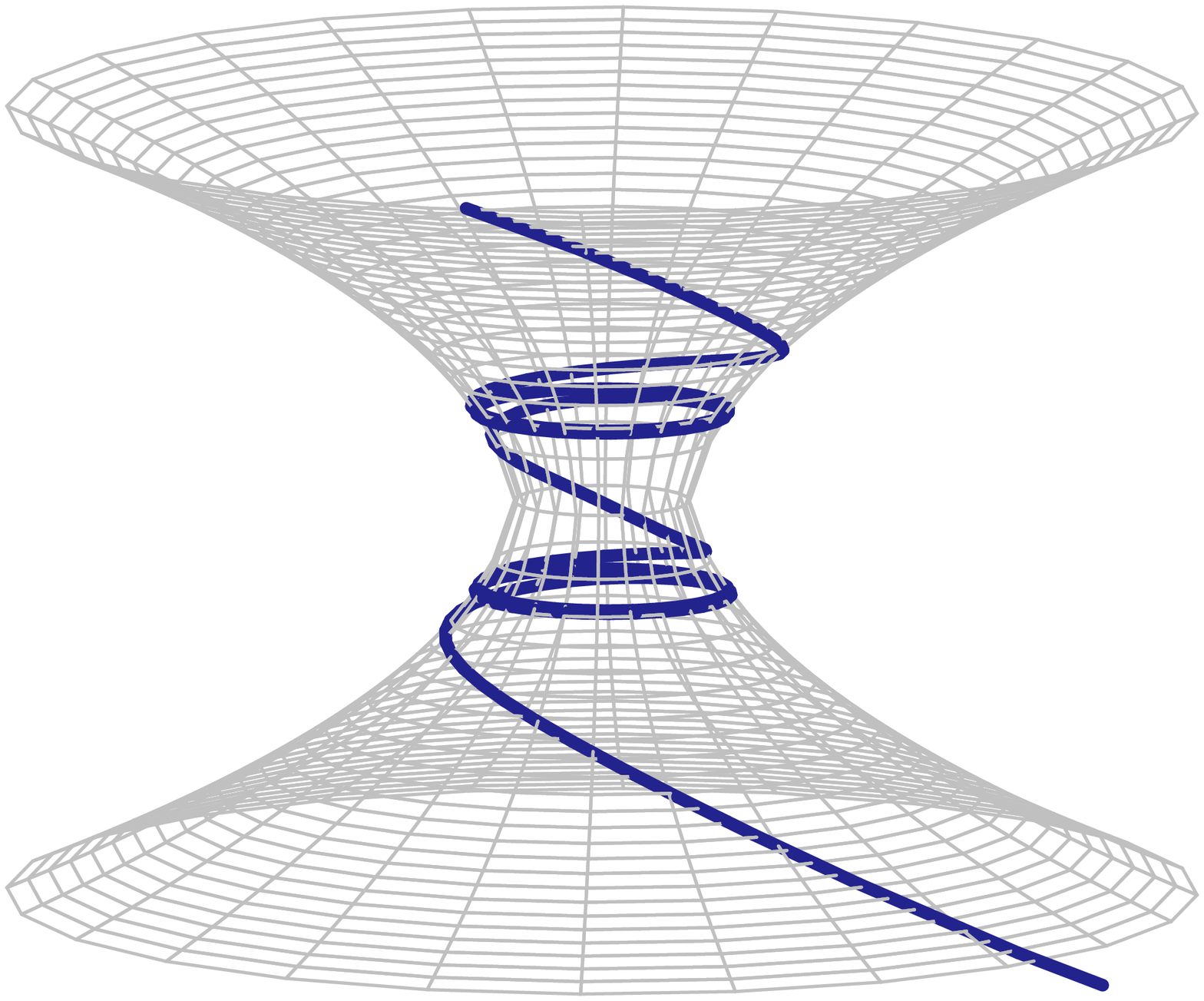}} \\
\subfigure[][$E=0.975, L=2.1$. BO.]{\label{kerr_orb4}\includegraphics[width=4.5cm]{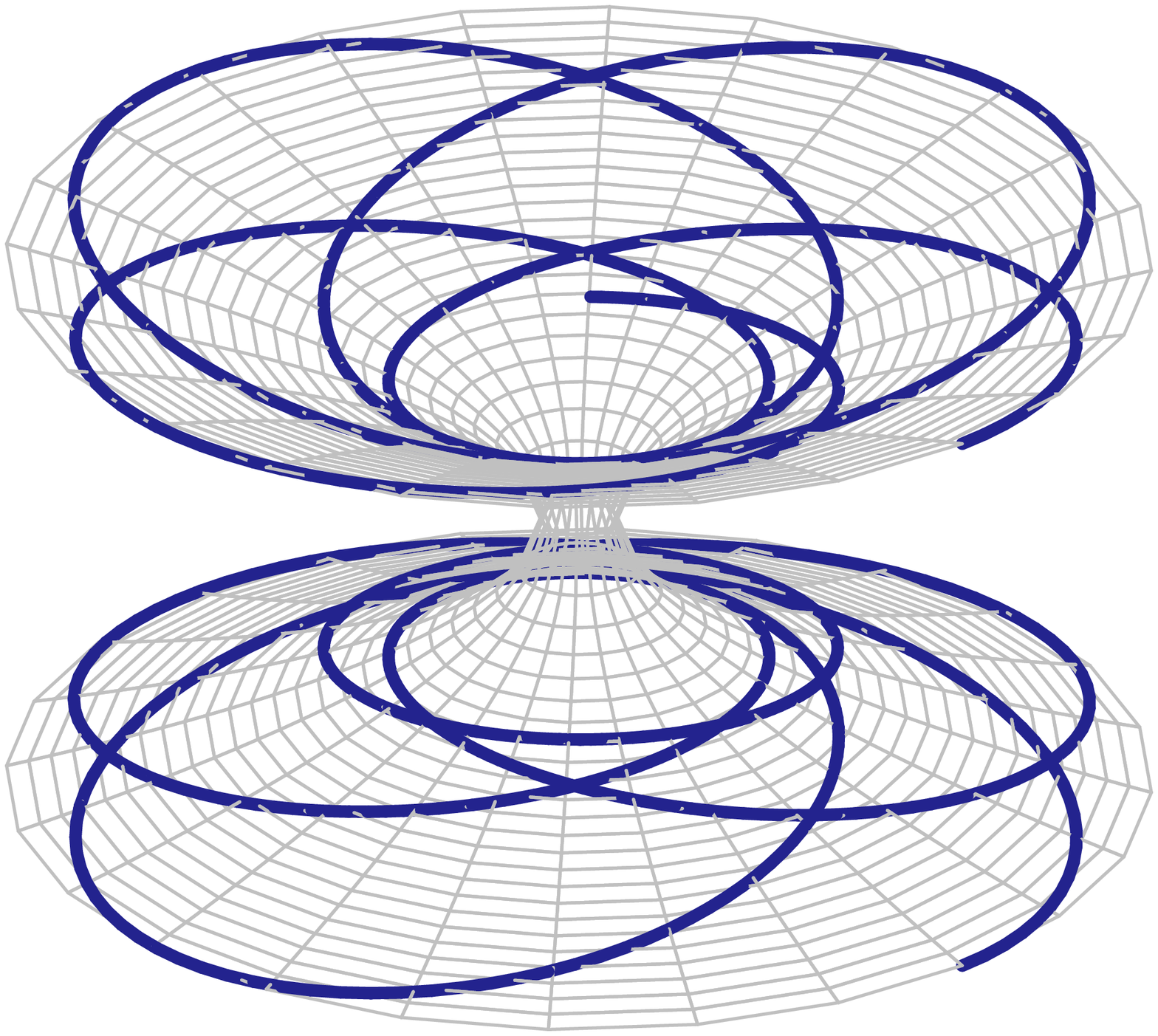}}
\subfigure[][$E=0.975, L=2.1$. TWB.]{\label{kerr_orb5}\includegraphics[width=4.5cm]{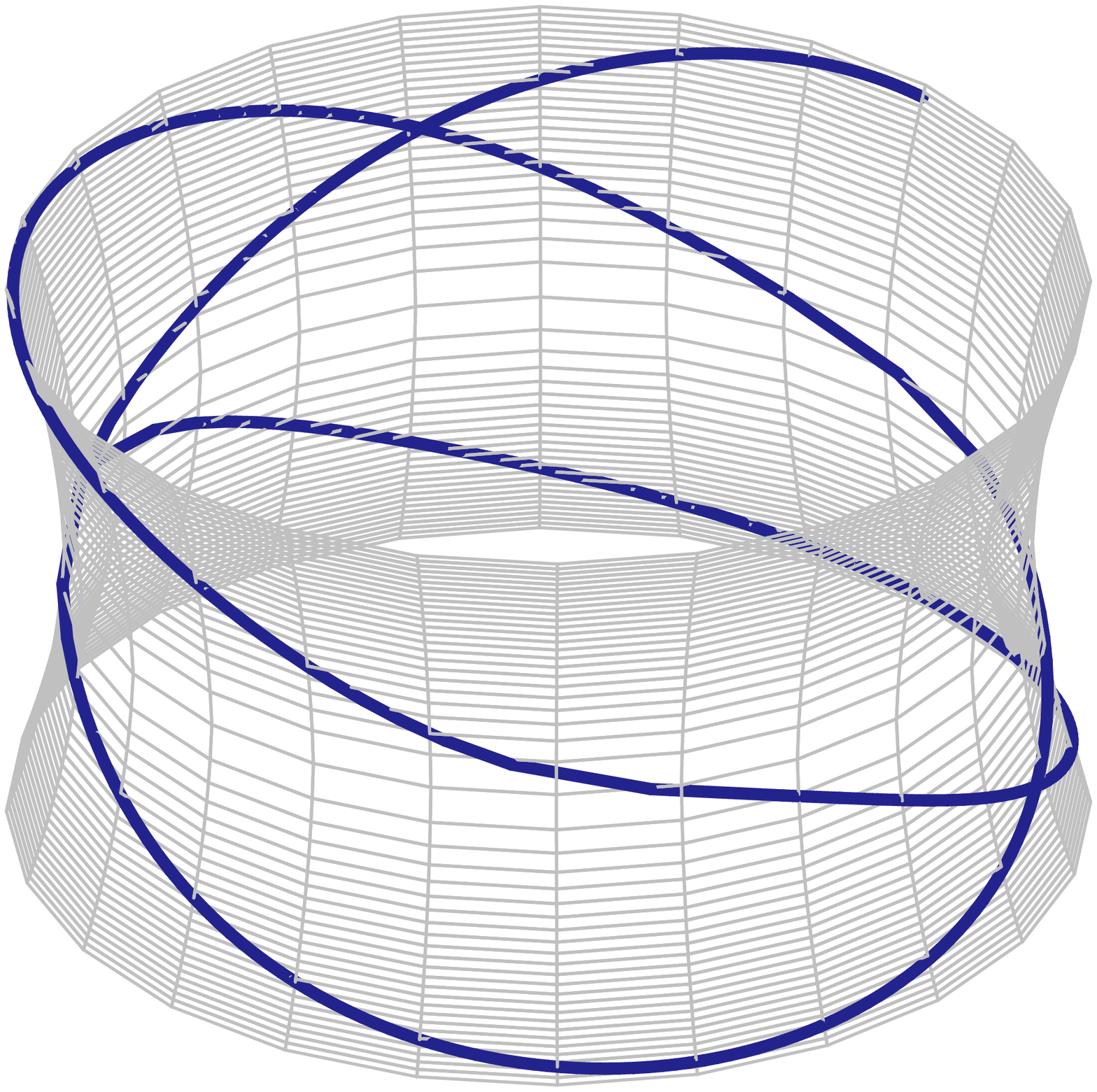}}
\end{center}
\caption{Orbits for a test particle with $\delta=1$ for different values of angular momentum and energy in the thin-shell Kerr wormhole spacetime. The two-world-bound (TWB) and two-world-escape (TWE) orbits stretch across the two regions of a Universe connected by a wormhole with throat parameter $b_0=0.99$ in~(\subref{kerr_orb1}-\subref{kerr_orb3}) and $b_0=0.96$ in~(\subref{kerr_orb4}-\subref{kerr_orb5}). Kerr--paramerer $a=0.2$. \label{fig:kerrorb}}
\end{figure*}

\begin{figure*}[th!]
\begin{center}
\subfigure[][$E=0.96492, L=-2.1$. TWB.]{\label{kerr_orb1_ergo_xy}\includegraphics[width=4.5cm]{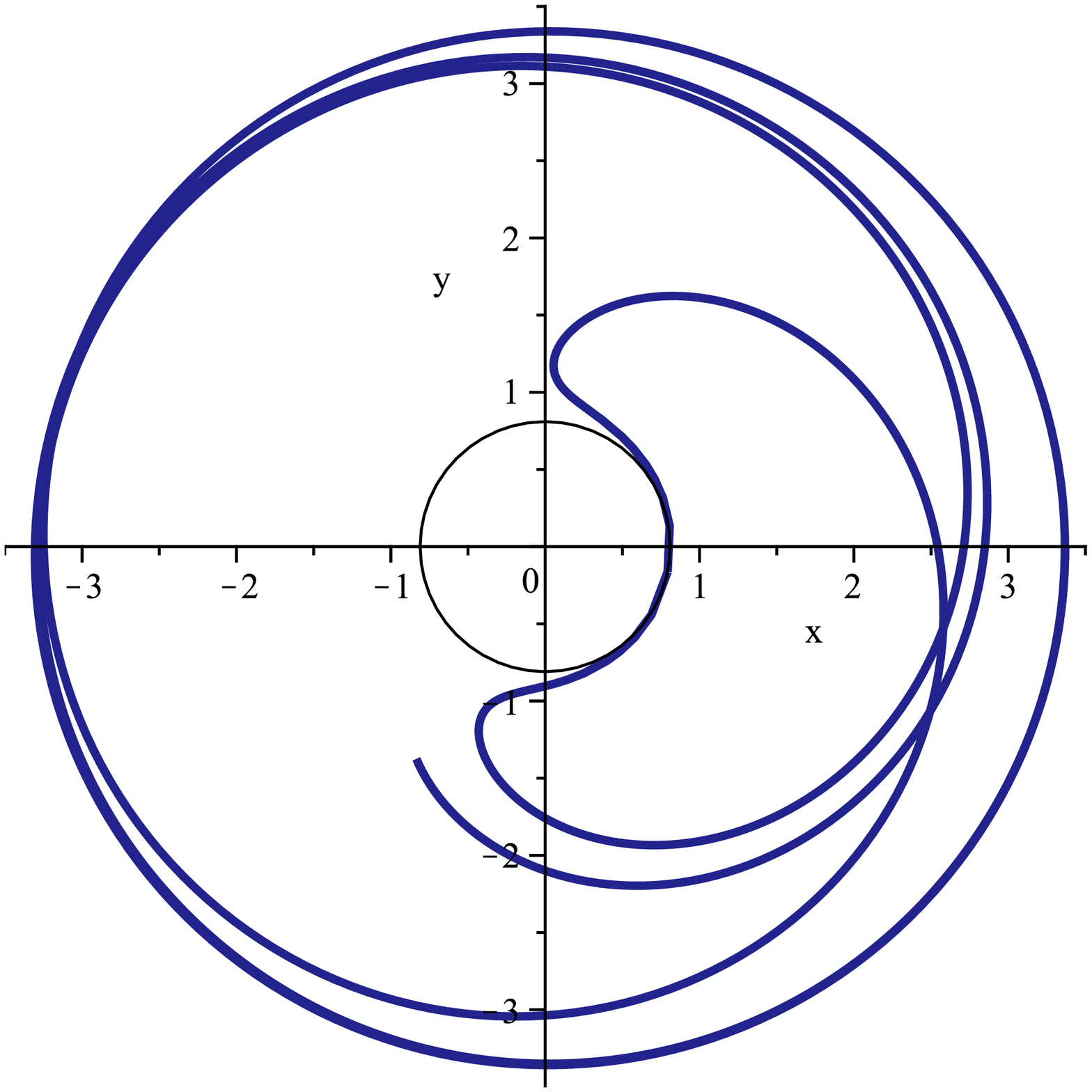}}
\subfigure[][$E=0.96492, L=-2.1$. TWB.]{\label{kerr_orb1_ergo}\includegraphics[width=4.5cm]{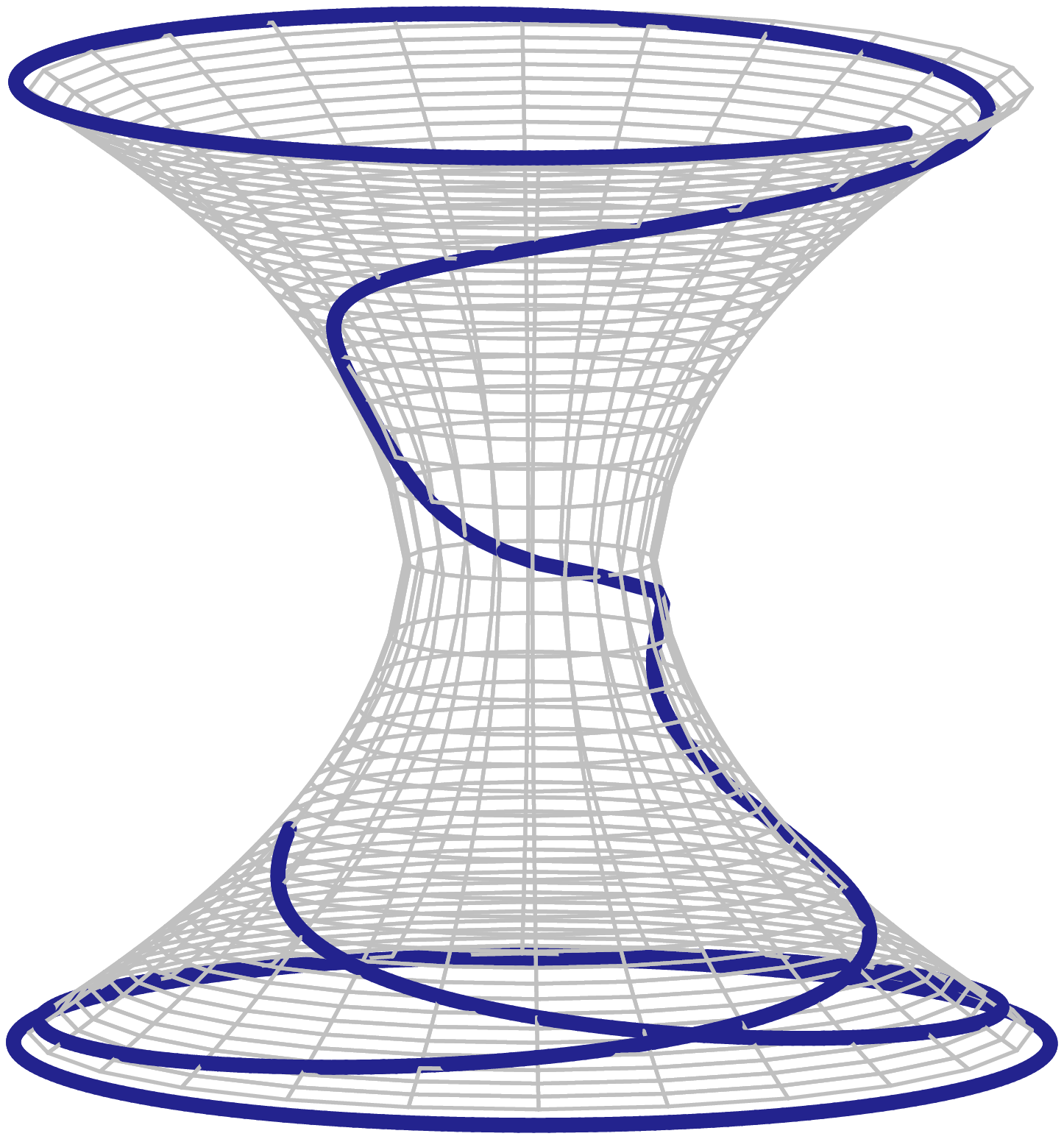}}
\subfigure[][$E=0.96492, L=-2.1$. EO.]{\label{kerr_orb2_ergo}\includegraphics[width=4.5cm]{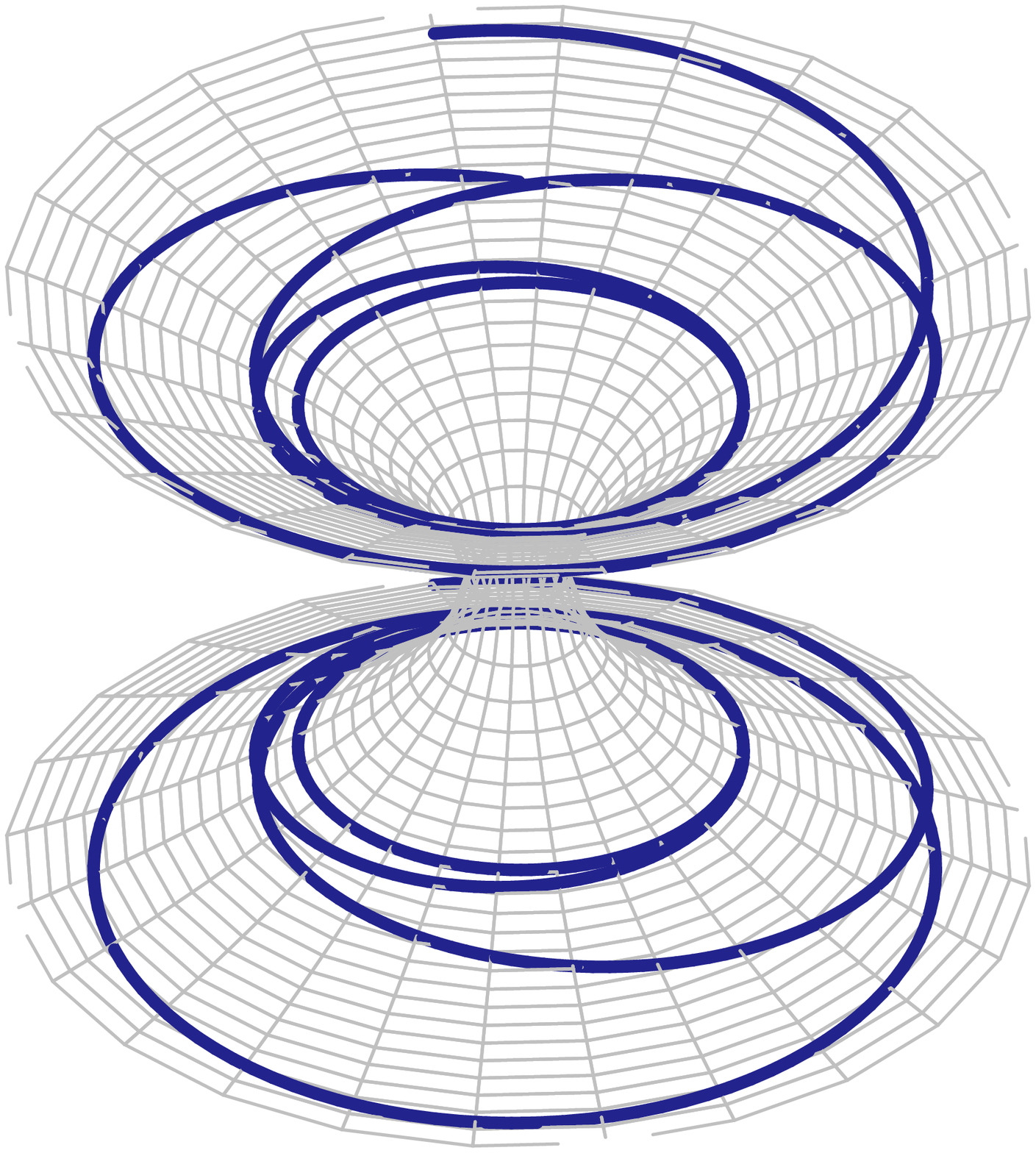}} \\
\subfigure[][$E=1.01, L=-2.1$. TWE.]{\label{kerr_orb3_ergo_xy}\includegraphics[width=4.5cm]{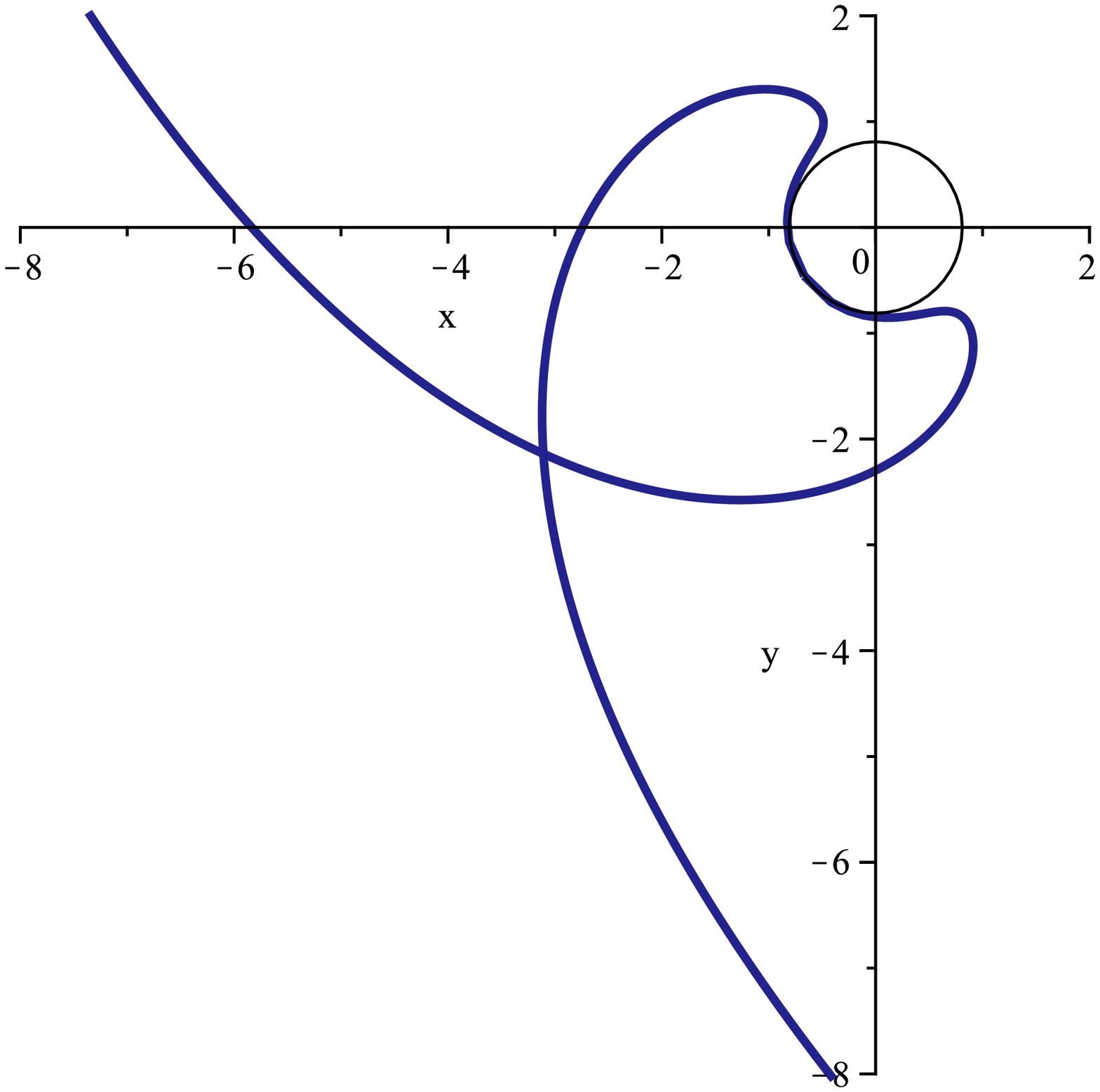}}
\subfigure[][$E=1.01, L=-2.1$. TWE.]{\label{kerr_orb3_ergo}\includegraphics[width=4.5cm]{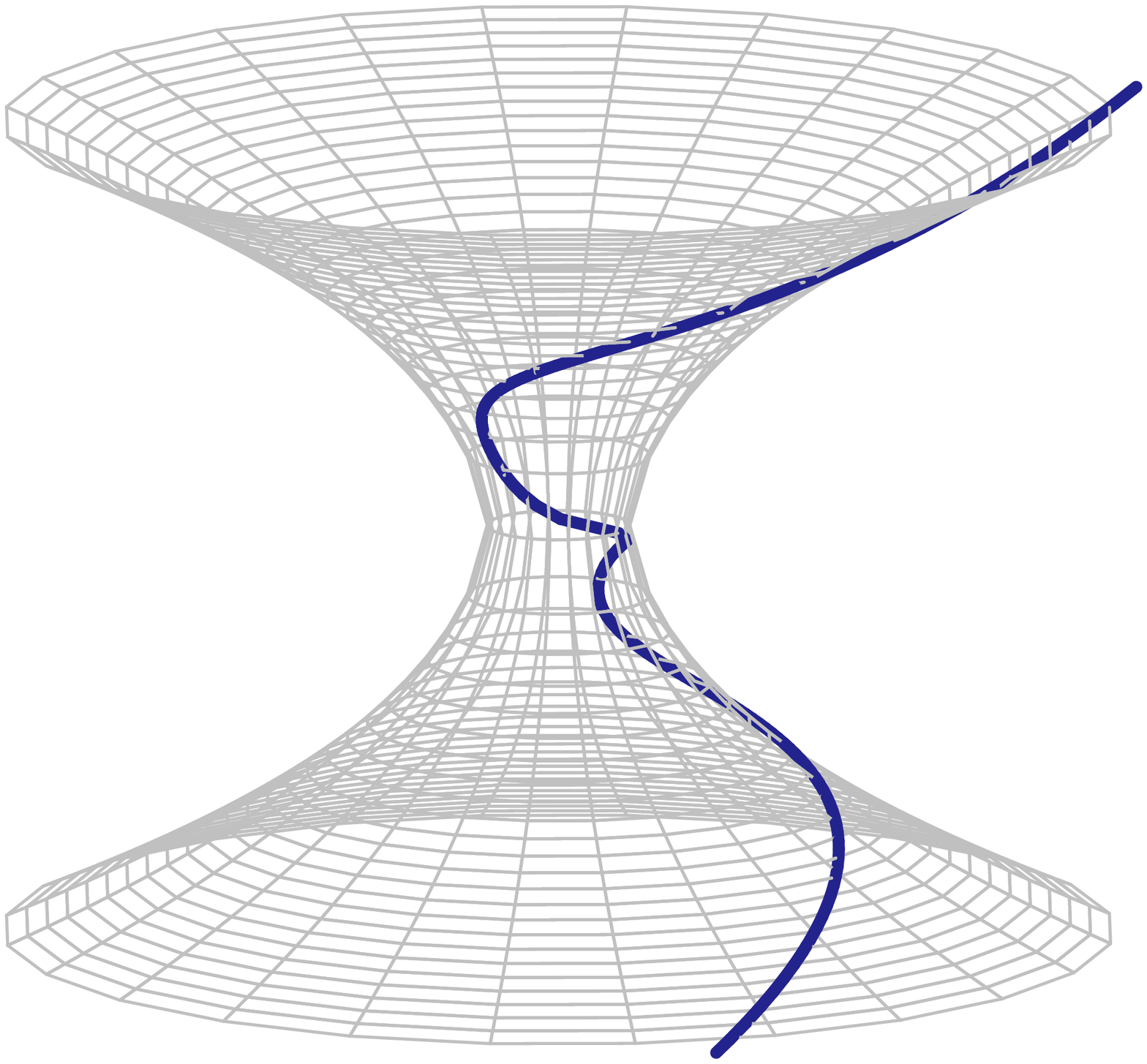}}
\end{center}
\caption{Orbits for a test particle with $\delta=1$ for different values of angular momentum and energy in the thin-shell Kerr wormhole spacetime. Kerr--paramerer $a=0.4$. The orbit~\subref{kerr_orb1_ergo_xy} is the $x$-$y$ projection of the orbit~\subref{kerr_orb1_ergo}. The black circle is the throat $b_0=0.81$. A test particle with the negative value of the angular momentum is dragged in the direction of the Kerr-wormhole rotation. This happens in the vicinity of the ergoregion which surrounds the throat of the wormhole. The third orbit~\subref{kerr_orb2_ergo} is a bound orbit for the same  $E$- and $L$-values. The described dragging can be also seen in the orbit shown in~\subref{kerr_orb3_ergo} and its $x$-$y$ projection~\subref{kerr_orb3_ergo_xy}. \label{fig:kerrorb_ergo}}
\end{figure*}

We illustrate these orbit types in Figs.\ref{fig:kerrorb} and~\ref{fig:kerrorb_ergo}. Fig.~\ref{fig:kerrorb_ergo}, plotted for opposite signs of the rotation parameter of the wormhole and the angular momentum of a test particle, also shows the influence of the ergosphere. It can be recognized in the directional change taking place shortly before the black circle is approached,
which symbolizes the wormhole throat in the pictures~\subref{kerr_orb1_ergo_xy} for the TWB and~\subref{kerr_orb3_ergo_xy} for the TWE orbit. The ergoregion in the Kerr wormhole spacetime surrounds the throat. The corresponding orbits are shown in~\subref{kerr_orb1_ergo} and~\subref{kerr_orb3_ergo}. The trajectories are embedded into the three-dimensional space with the coordinates $(x,y,z)$, where $x=r\cos(\varphi)$, $y=r\sin(\varphi)$, and $r$, $\varphi$ and $z$ are given by~\eqref{eq:kerr-r3},~\eqref{eq:kerr-phi4} and~\eqref{kerr_emb3} correspondingly.

\bigskip
\paragraph{Geodesics for $\delta=0$.}

\begin{figure*}[th!]
\begin{center}
\subfigure[][$E=1.271, L=1.5$. TWB.]{\label{kerr_orb1_light}\includegraphics[width=4.5cm]{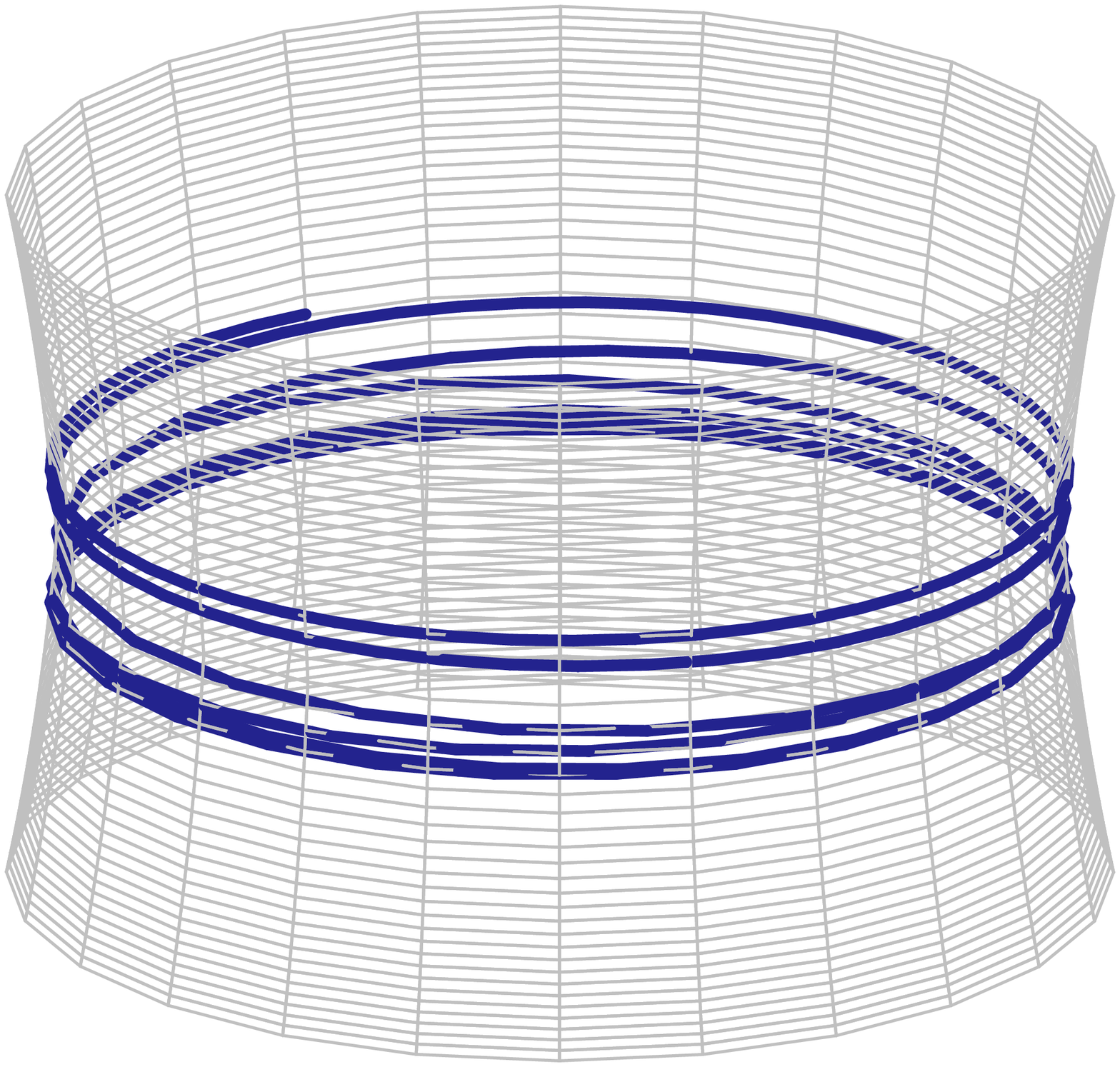}}
\subfigure[][$E=1.271, L=1.5$. EO.]{\label{kerr_orb2_light}\includegraphics[width=4.5cm]{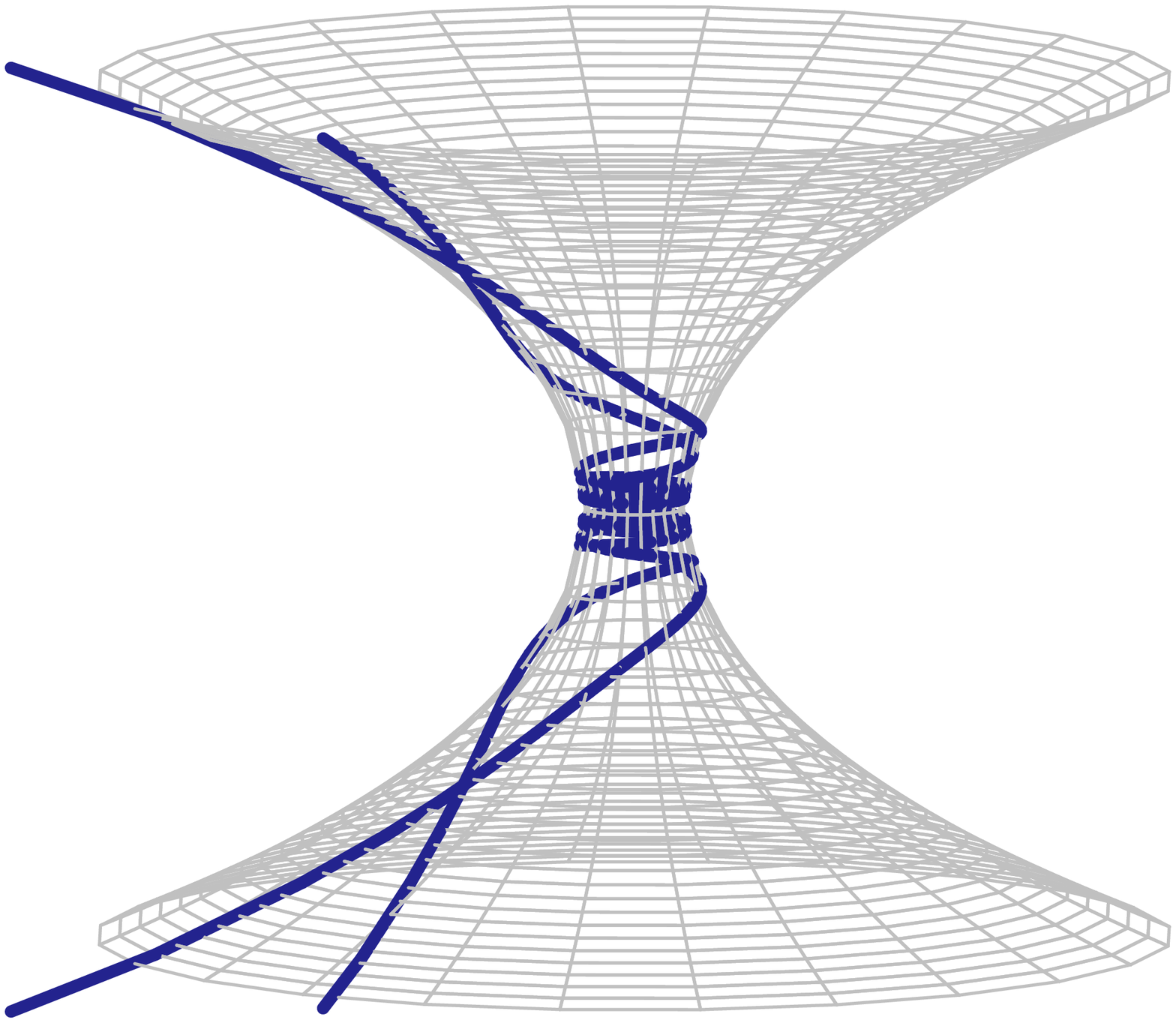}}
\subfigure[][$E=1.272, L=1.5$. TWE.]{\label{kerr_orb3_light}\includegraphics[width=7.5cm]{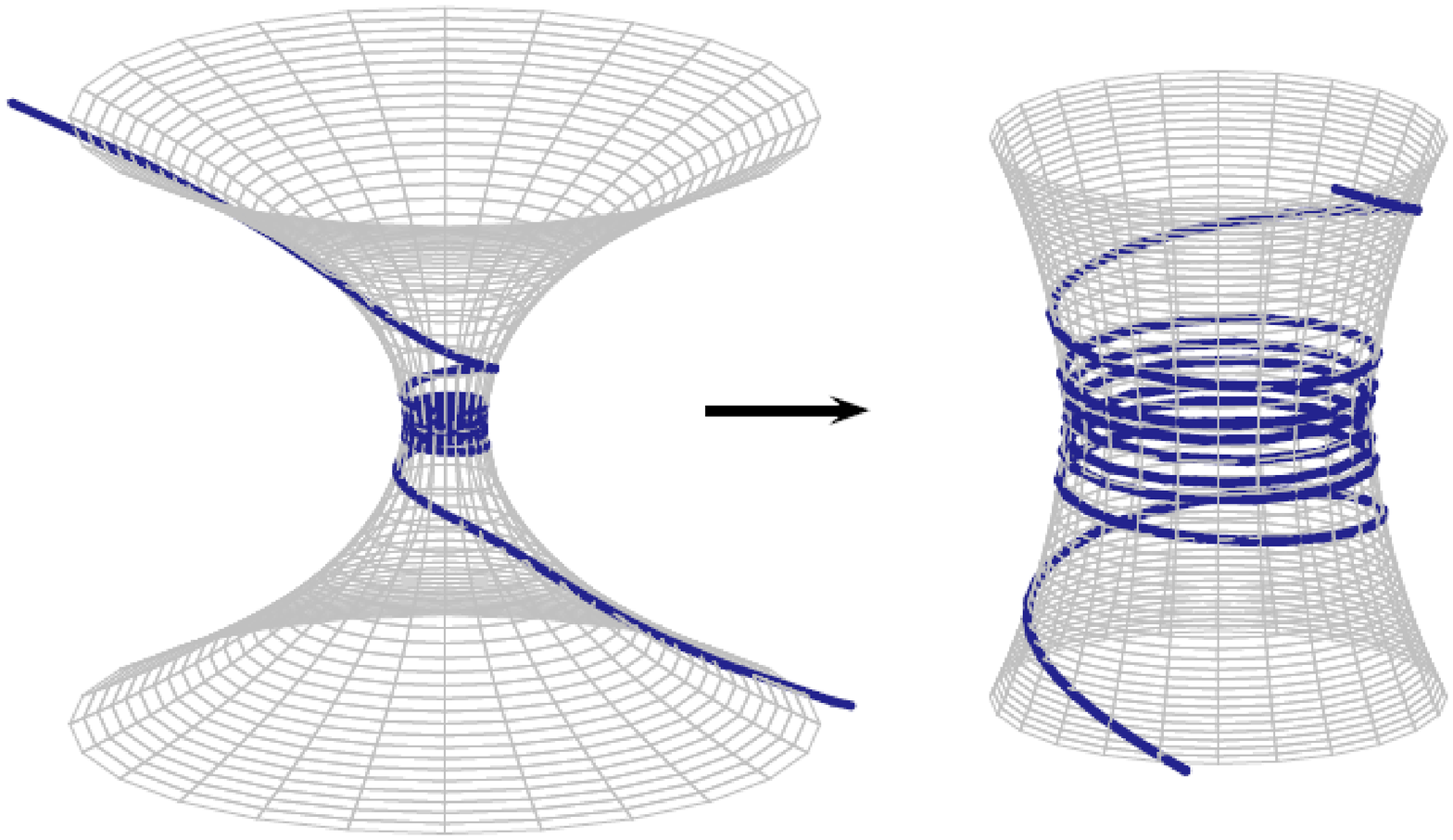}}
\end{center}
\caption{Orbits for a massless test particle ($\delta=0$) in the thin-shell Kerr wormhole spacetime. Kerr--paramerer $a=0.49$. Plot\subref{kerr_orb1_light} shows a two-world bound orbit and plot\subref{kerr_orb2_light} two escape orbits in the upper and lower parts of the Universe. Plot\subref{kerr_orb3_light} shows a test particle moving in the upper and lower parts of the Universe on a two-world escape orbit. Here $b_0=0.61$.  \label{fig:kerrorb_light}}
\end{figure*}

\begin{figure*}[th!]
\begin{center}
\subfigure[][$E=0.4306, L=-1.5$. TWB.]{\label{kerr_orb1_light_ergo_xy}\includegraphics[width=4.5cm]{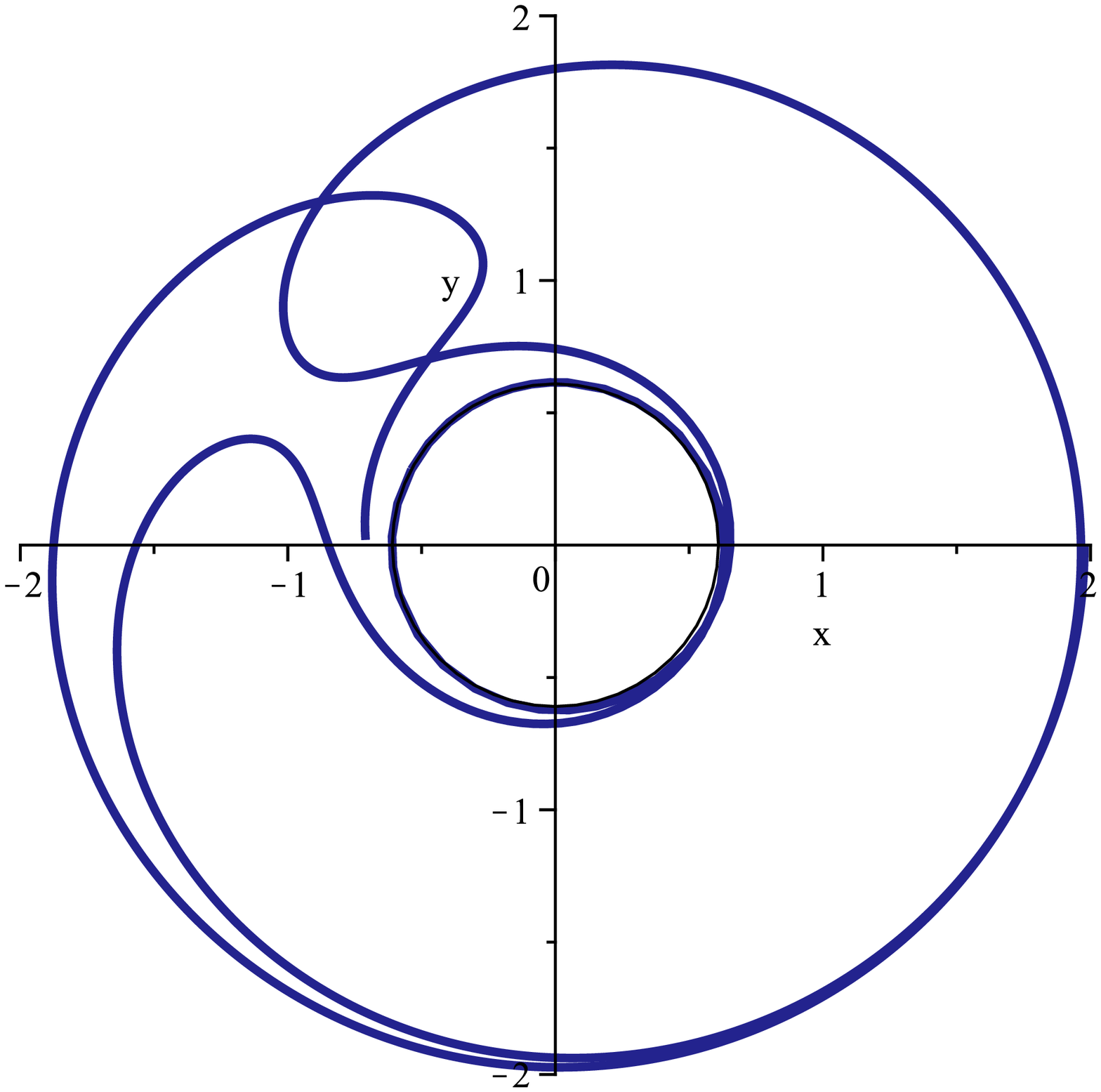}}
\subfigure[][$E=0.4306, L=-1.5$. TWB.]{\label{kerr_orb1_light_ergo}\includegraphics[width=3.1cm]{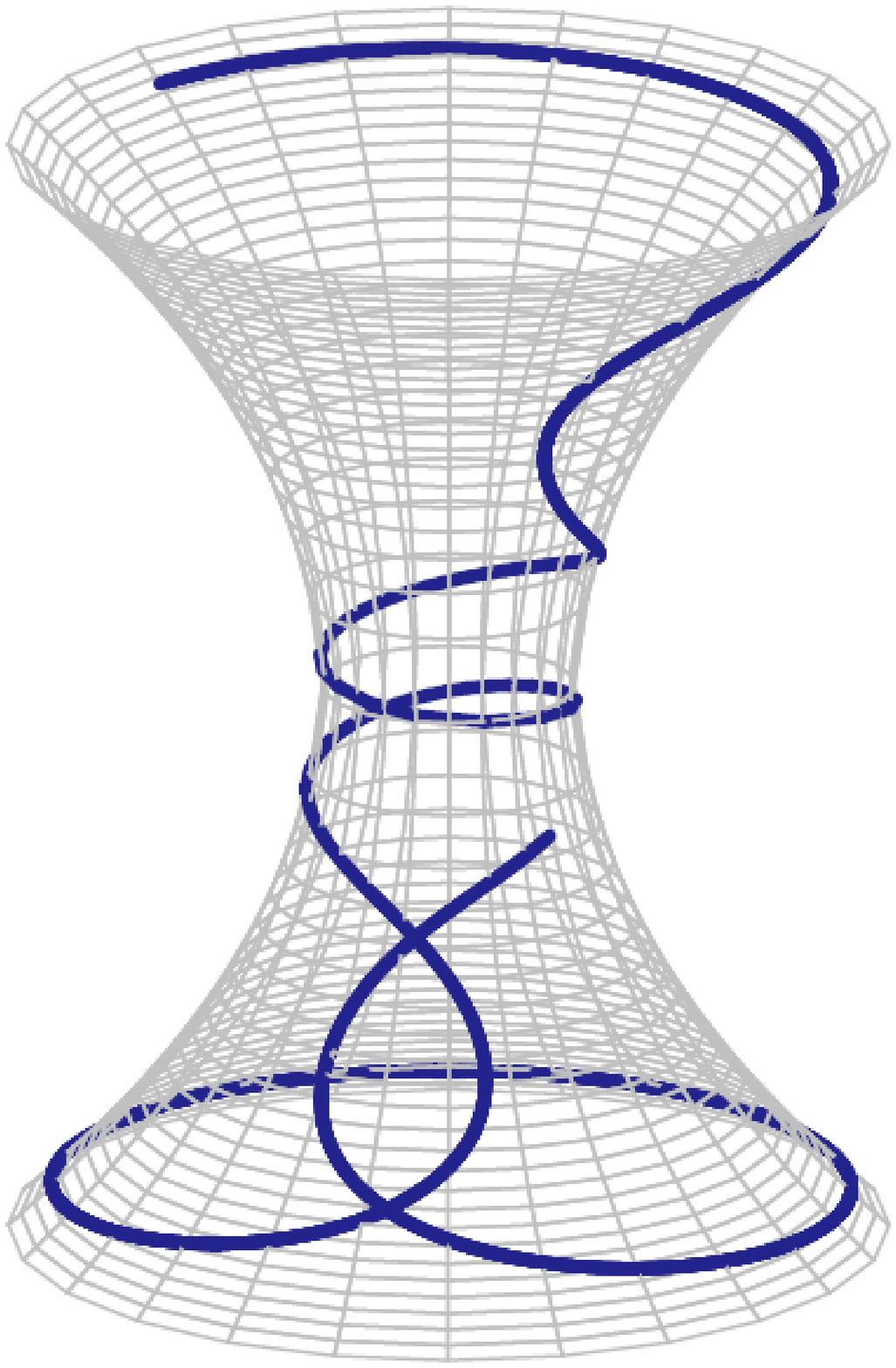}}
\subfigure[][$E=0.4306, L=-1.5$. EO.]{\label{kerr_orb2_light_ergo}\includegraphics[width=4.5cm]{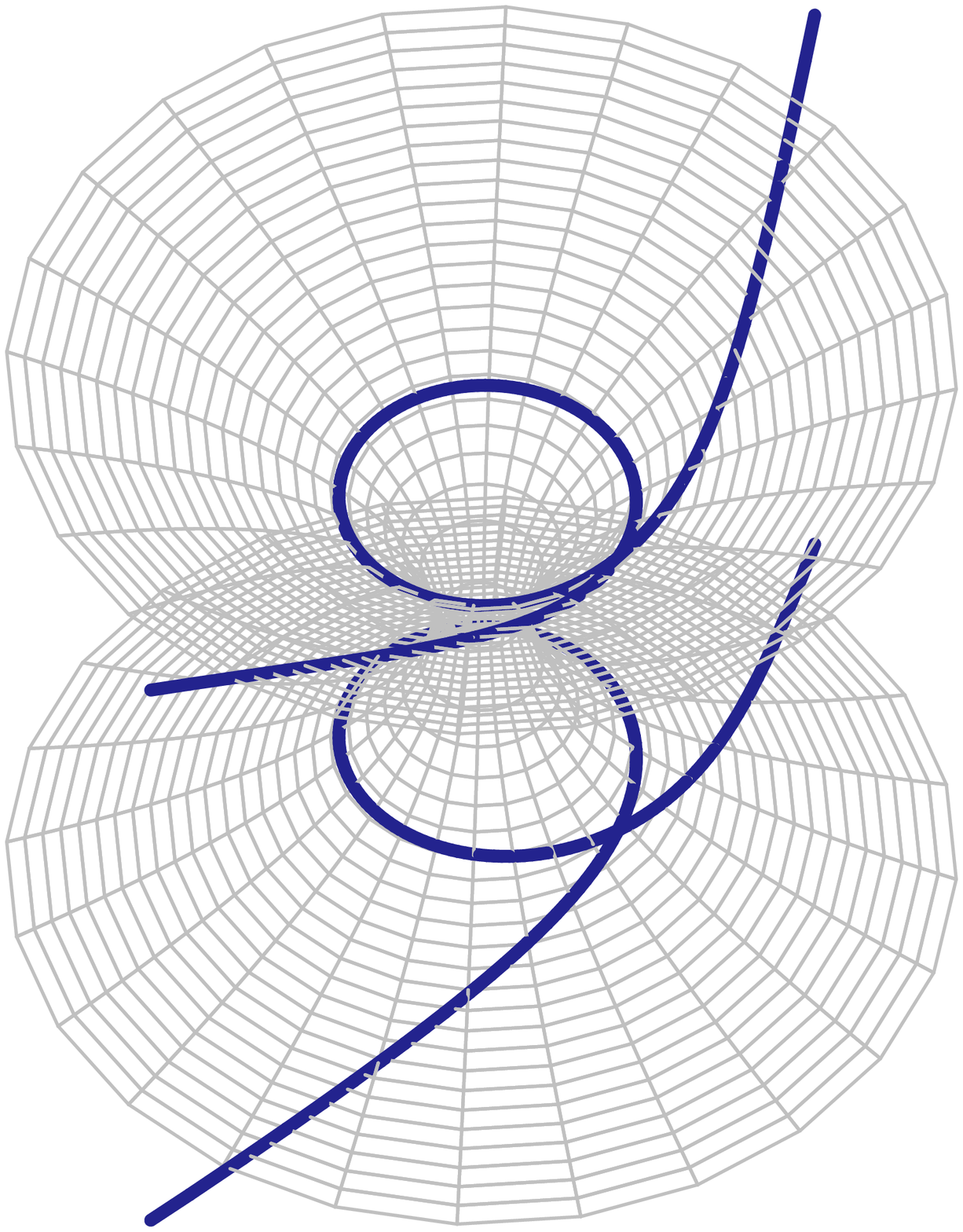}} \\
\subfigure[][$E=0.4307, L=-1.5$. TWE.]{\label{kerr_orb3_light_ergo_xy}\includegraphics[width=4.5cm]{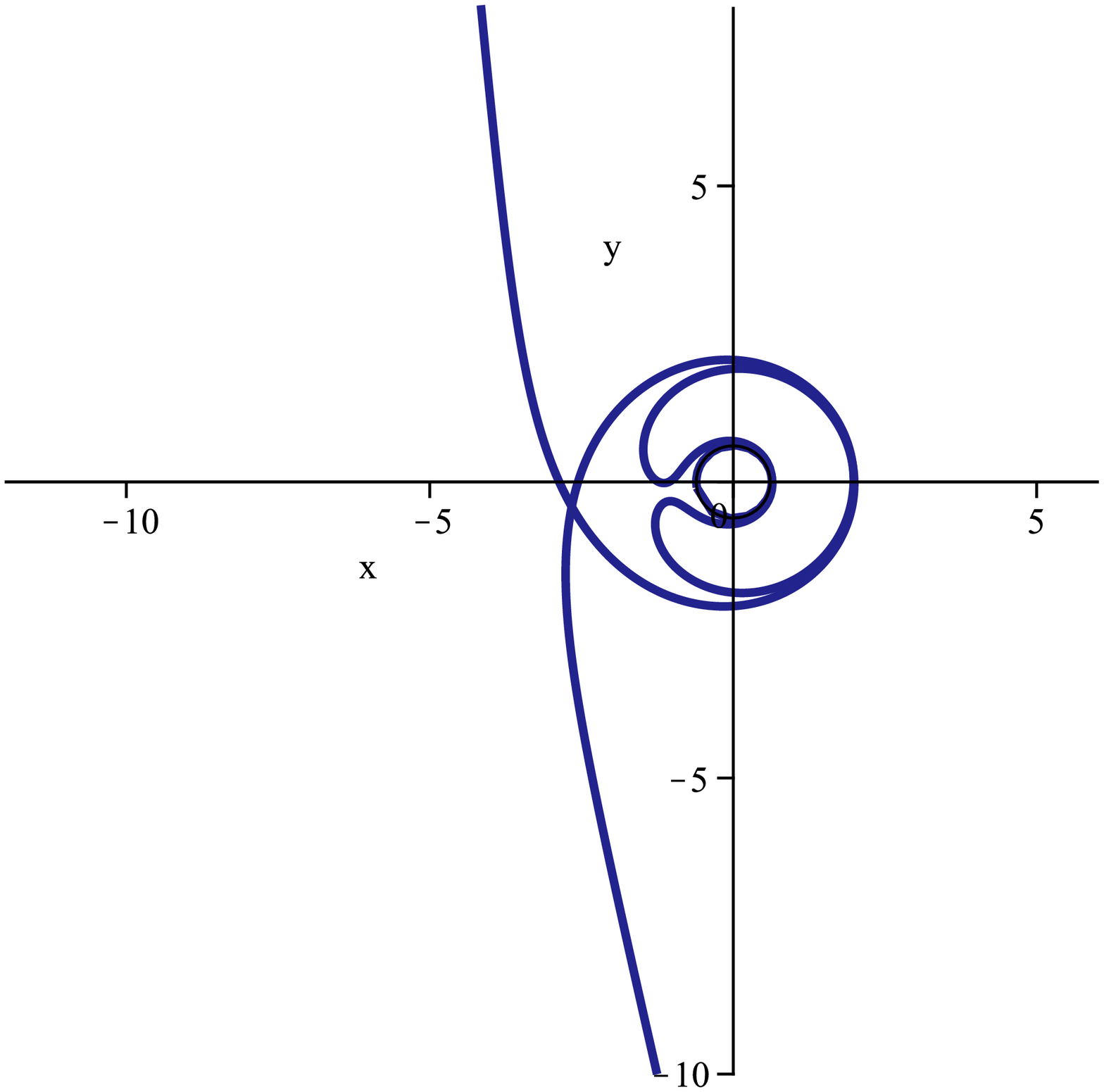}}
\subfigure[][$E=0.4307, L=-1.5$. TWE.]{\label{kerr_orb3_light_ergo}\includegraphics[width=4.5cm]{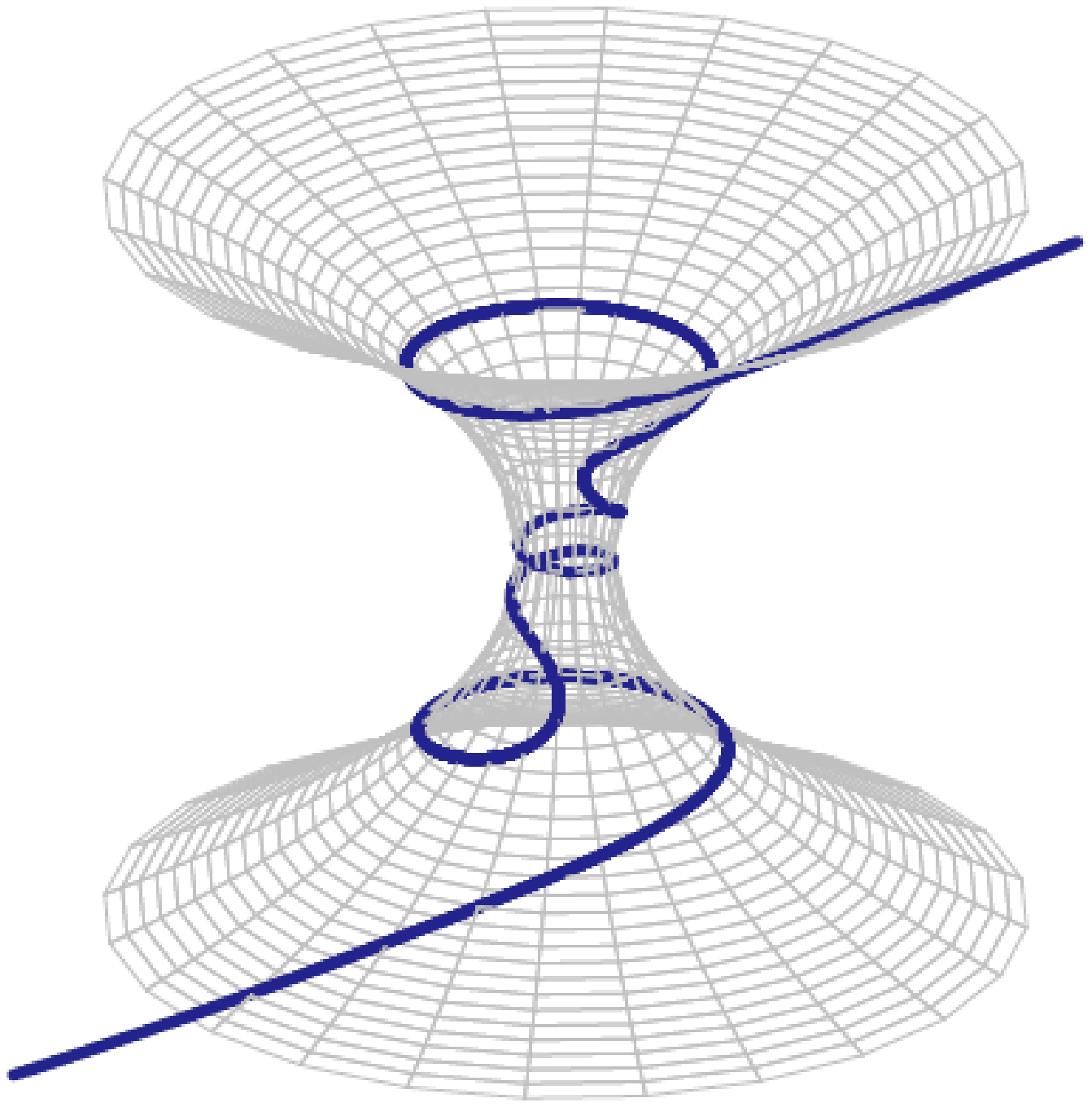}}
\end{center}
\caption{Orbits in the thin-shell Kerr wormhole spacetime for a massless test particle ($\delta=0$) when a particle and a wormhole are counterrotating. Kerr--parameter $a=0.49$. Plot\subref{kerr_orb1_light_ergo} shows a two-world bound orbit and plot\subref{kerr_orb2_light_ergo} two escape orbits in the upper and lower parts of the Universe. The picture\subref{kerr_orb1_light_ergo_xy} is a projection of~\subref{kerr_orb1_light_ergo} onto the $x$-$y$ plane. Plots\subref{kerr_orb3_light_ergo_xy} and~\subref{kerr_orb3_light_ergo} show a two-world escape orbit extending over upper and lower parts of the Universe. Here $b_0=0.609$.  \label{fig:kerrorb_light_ergo}}
\end{figure*}

Like in the Kerr black hole spacetime no planetary bound orbits exist in the Kerr thin-shell wormhole spacetime, and only the orbits of type $A$ (two-world escape TWE), $B$ (two-world bound TWB) and $C$ (two-world bound TWB and escape EO) can be found. Fig.\ref{fig:kerrorb_light} visualizes some orbits for corotating massless test particles and Fig.\ref{fig:kerrorb_light_ergo} for counterrotating ones. In the last case a particle is dragged into the direction of wormhole rotation in the vicinity of ergosphere which is best seen in the projection (figure~\ref{kerr_orb1_light_ergo_xy}) for the embedded two-world bound orbit (figure~\subref{kerr_orb1_light_ergo}), and the projection (figure~\ref{kerr_orb3_light_ergo_xy}) for the embedded two-world escape orbit~\subref{kerr_orb3_light_ergo}.

\vspace*{0.2cm}
\section{Conclusion}

We have studied the motion of massive and  massless test particles in the Schwarzschild and Kerr thin-shell wormhole spacetimes. These traversable wormhole spacetimes were constructed by the cut-and-paste method and represent geodesically complete manifolds. The solution of the geodesic equations is given analytically in terms of elliptic Jacobi and Weierstrass functions. We have shown that in the {\em traversable} Schwarzschild and Kerr wormhole spacetimes bound and escape orbits connecting the upper and lower parts of the Universe exist. Planetary bound orbits exist only for not too wide a throat of the wormhole in both cases. If the throat is sufficiently narrow (i.e. the throat parameter is chosen to be smaller than the ergosphere), a test particle is dragged into the direction of the wormhole rotation in the vicinity of the ergosphere. Here extraction of energy due to the Penrose process is possible. 

Free particles moving on geodesics in the thin-shell wormhole spacetimes considered here encounter no exotic matter needed to maintain the wormhole and feel no tidal forces so that they will simply be transferred into the other universe. This possibility for travelers to avoid regions of exotic material in their traversal of the wormhole was discussed by Visser~\cite{Visser1,Visser2} and Teo~\cite{Teo}. 

We will discuss the general geodesics - without restriction to the equatorial plane - for the travesable Kerr thin-shell wormhole spacetime in~\cite{KagramanovaSmolarek2} where the solution of the geodesic equations is also given by elliptic $\wp$, $\zeta$ and $\sigma$-functions.

\section*{Acknowledgement}

We would like to thank Jutta Kunz and Burkhard Kleihaus for helpful discussions. 
We also gratefully acknowledge financial support of the German Research Foundation DFG and the support within the framework of the DFG Research Training Group 1620 {\it Models of gravity}.

\bibliographystyle{unsrt}

\end{document}